\documentclass[fleqn,usenatbib]{mnras}

\usepackage{newtxtext,newtxmath}
\usepackage{bm}
\usepackage{comment} 

\usepackage[T1]{fontenc}

\DeclareRobustCommand{\VAN}[3]{#2}
\let\VANthebibliography\thebibliography
\def\thebibliography{\DeclareRobustCommand{\VAN}[3]{##3}\VANthebibliography}

\usepackage{graphicx}	
\usepackage{ulem}
\usepackage{physics}
\usepackage{siunitx}
\usepackage[capitalise]{cleveref}
\usepackage{hyperref}
\usepackage{lineno}

\usepackage[dvipsnames]{xcolor}

\title[Likelihoods for cluster abundance cosmology]{Testing the accuracy of likelihoods for cluster abundance cosmology}

\author[Payerne, Murray, Combet et al.]{
\parbox{\textwidth}{
\Large
C. Payerne,$^{1}$\thanks{E-mail: constantin.payerne@lpsc.in2p3.fr}
C. Murray,$^{1}$
C. Combet,$^{1}$
C. Doux,$^{1}$
A. Fumagalli,$^{2,3,4,5}$ 
M. Penna-Lima$^{6}$
}
\vspace{0.4cm}
\\
\parbox{\textwidth}{
$^{1}$ Université Grenoble Alpes, CNRS, LPSC-IN2P3, 38000 Grenoble, France\\
$^{2}$ Dipartimento di Fisica - Sezione di Astronomia, Universit\'a di Trieste, Via Tiepolo 11, I-34131 Trieste, Italy\\
$^{3}$ INAF-Osservatorio Astronomico di Trieste, Via G. B. Tiepolo 11, I-34131 Trieste, Italy\\
$^{4}$ IFPU, Institute for Fundamental Physics of the Universe, via Beirut 2, 34151 Trieste, Italy\\
$^{5}$ INFN, Sezione di Trieste, Via Valerio 2, I-34127 Trieste TS, Italy\\
$^{6}$ Instituto de Física, Universidade de Brasília, 70910-900, Brasília, DF, Brazil
}
}

\date{Accepted XXX. Received YYY; in original form ZZZ}

\pubyear{2015}

\begin{document}
\label{firstpage}
\pagerange{\pageref{firstpage}--\pageref{lastpage}}
\maketitle
\begin{abstract}
The abundance of galaxy clusters is a sensitive probe to the amplitude of matter density fluctuations, the total amount of matter in the Universe as well as its expansion history. Inferring correct values and accurate uncertainties of cosmological parameters requires accurate knowledge of cluster abundance statistics, encoded in the likelihood function. In this paper, we test the accuracy of cluster abundance likelihoods used in the literature, namely the Poisson and Gaussian likelihoods as well as the more complete description of the Gauss-Poisson Compound likelihood. This is repeated for a variety of binning choices and analysis setups. In order to evaluate the accuracy of a given likelihood, this work compares individual posterior covariances to the covariance of estimators over the 1000 simulated dark matter halo catalogs obtained from PINOCCHIO algorithm. We find that for Rubin/LSST and Euclid-like surveys the Gaussian likelihood gives robust constraints over a large range of binning choices. The Poisson likelihood, that does not account for sample covariance, always underestimates the errors on the parameters, even when the sample volume is reduced or only high-mass clusters are considered. We find no benefit in using the more complex Gauss-Poisson Compound likelihood as it gives essentially the same results as the Gaussian likelihood, but at a greater computational cost. Finally, in this ideal setup, we note only a small gain on the parameter error bars when using a large number of bins in the mass-redshift plane.
\end{abstract}

\begin{keywords}
galaxies: clusters: general --
cosmology: cosmological parameters -- 
methods: statistical
\end{keywords}



\section{Introduction}
\label{section:introduction}

Galaxy clusters have been essential in the construction of the standard model of cosmology, providing some of the first evidence of dark matter \citep{zwicky} through the motions of galaxies within galaxy clusters and their spatial distribution provided evidence for the primordial origin of density fluctuations \citep{kaiser1984spatial}. Since their formation is sensitively dependent on the properties of both dark energy and dark matter, future measurements of the masses of galaxy clusters and their abundance will hopefully deliver deep insights on the nature of these elusive components of the Universe. 

The primary cosmological probe for galaxy clusters is the counting of their abundance as a function of mass and redshift. The Rubin Observatory Legacy Survey of Space and Time \citep{LSST} and the \textit{Euclid} mission survey \citep{laureijs2011euclid} will mark a new era in the study of galaxy clusters for cosmology due to an order of magnitude increase of the number of galaxy clusters with precision weak gravitational lensing information. In order to fully take advantage of these datasets it is important to have an accurate statistical description of the abundance of clusters in the Universe.

Galaxy clusters form in the densest regions of the cosmic web. Their formation is a complex, non-linear process, dominated by gravitational forces although it is important also to consider baryonic effects \citep{henson2016impact}. They form from the collapse of overdense regions of the initially Gaussian matter density field of the early Universe. Early attempts at describing their statistics were with the excursion-set model \citep{press1974formation,bond1991excursion,sheth_tormen}. Such models must simplify greatly the complexity of cluster formation; assuming spherical or ellipsoidal collapse of over-dense regions. Therefore to understand in more detail the distribution of clusters in mass and redshift cosmological simulations are essential \citep{jenkins2001,evrard2002,tinker2008}.

Much of this existing work however has focused on the prediction of the mean of the cluster abundance for a given mass and redshift, rather than the distribution of the number of clusters at a given mass and redshift. Essentially this is the challenge of this study.

In practice three cluster likelihoods have been used to perform cluster count cosmology: a Poissonian likelihood (e.g., \citealp{mantz2015weighing,bocquet2019cluster,ade2016planck}), a Gaussian likelihood (e.g., \citealp{abbott2020dark,2022arXiv220307398L}) and, more recently, a combination of the two \citep{lesci2022amico}. Each of these likelihoods are described in detail later in the article.

The latent likelihood of the distribution of cluster abundances in the Universe is not Poissonian, Gaussian or a combination of the two. By the latent likelihood here we refer to the unknown underlying likelihood which would perfectly describe the distribution of galaxy clusters in the Universe given a cosmological model. The halo model of structures in the Universe (in which each of these likelihoods is to some extent built upon) is a model, which at some level will break down. The distribution of galaxy clusters in a universe predicted by a cosmological model will be more complicated than any of these likelihoods. Therefore the assumption is that the cluster abundance distribution is well described by these models. If not this will impact the constraint of cosmological parameters, possibly leading to mis-estimation of parameters errors and even bias the results.

This work aims to answer a very general question: is our statistical description of the abundance of clusters sufficiently accurate for future surveys? Within the cosmology community at large this has been assessed in the context of mis-estimation of the covariance matrices for cosmological analyses \citep{dodelson2013effect,taylor2014estimating,sellentin2016quantifying}. Here we move beyond the Gaussian approximation in the specific case of cluster abundance cosmology, to answer to what extent are the Poissonian and Gaussian models insufficient. The insufficiency of Gaussian likelihoods in the case of the CFHTLenS cosmic shear survey was considered by \cite{sellentin2018insufficiency}, where they note since that one-point statistic of the correlations are skewed, the Gaussian approximation is not necessarily accurate. 

In addition, we seek to answer multiple related questions within this work. They are: (i) which likelihoods can sufficiently describe the abundance of galaxy clusters as a function of mass and redshift and provide robust constraints, (ii) among those, does a given likelihood provides stronger cosmological constraints than the others?, and (iii) is there an optimal binning regime that allows us to extract the most cosmological information?

To answer these questions we present a comparison of the cluster distributions using a set of cosmological simulations and compare each theoretical model to that measured in the distribution. Subsequently, we show the effects on the obtained cosmological parameters and the estimated posterior of the cosmological parameters.

In \cref{section:cluster_distributions}, we give an overview of the three different likelihoods for cluster abundance cosmology used in this work. In \cref{section:likelihood_accuracy}, we present a framework for testing the accuracy of each of these likelihoods. The 1000 PINOCCHIO simulations that we use in this paper are presented in \cref{section:pinocchio}. \cref{sec:covariance} describes the covariance matrices used in the Gaussian and Gauss-Poisson Compound likelihoods and \cref{sec:methodology} presents the methodology of the work and the technical details of the posterior sampling. The main results of this articles are given in \cref{section:results}, in which using 1000 PINOCCHIO simulations we obtain the posteriors with each likelihood. By comparing the individual posteriors to the distribution of parameters estimated from the 1000 simulations we can test the accuracy and precision of each of our likelihoods. Finally, we discuss the conclusions of our work in \cref{sec:conclusions}.

\section{Likelihoods for cluster abundance cosmology}
\label{section:cluster_distributions}

In the following Section we describe the three different likelihoods used in this work; the Gauss-Poisson Compound likelihood, the Poissonian likelihood and the Gaussian likelihood.

\subsection{Gauss-Poisson Compound likelihood (GPC) likelihood}

Galaxy clusters trace the matter density field in the universe. The abundance, or number density, of galaxy clusters follows spatial fluctuations of the underlying matter density field. This relation can be parameterised by taking clusters to be biased tracers of the underlying density field \citep[see][for a review of the halo model]{cooray2002halo}. The overdensity of galaxy clusters, $\delta_h$, in a region centered at $\textbf{x}$,
\begin{equation}
    \delta_h(\mathbf{x})=b(m,z) \delta(\mathbf{x})\,,
    \label{eq:halo_overdensity}
\end{equation}
where $b(m,z)$ is the linear bias parameter which is a function of both cluster mass and redshift and $\delta(\mathbf{x},z)$ is the underlying matter density field. Using \cref{eq:halo_overdensity} the local halo number density, $n$, can be expressed as,

\begin{equation}
    n(m,z,\mathbf{x}) = \bar{n}(m,z)(1 + b(m,z) \delta(\mathbf{x},z))\,,
    \label{eq:lodal_density}
\end{equation}
where $\bar{n}$ is the predicted mean number density of clusters at a given mass and redshift for a given cosmological model. As we are concerned with fluctuations on the size of the survey, the matter density field $\delta(\mathbf{x},z)$ will be well described by a Gaussian random field; the same would not be true on smaller scales.

The abundance of galaxy clusters is the count of clusters in a specific spatial region, such as the observed number of galaxy clusters will be a Poisson realisation of the local halo number density, $n$, which in turn is a Gaussian realisation of $\bar{n}$ depending on the underlying density field of the survey region. 

\citet{2003_SV_HU} proposed a Gauss-Poisson Compound likelihood for cluster abundance cosmology that combines these two statistical effects; the average number of clusters $\lambda_{ij}$ within a given $i$-th mass and $j$-th redshift is calculated from the halo mass function $\frac{dn(m,z)}{dm}$ predicting the average number density of halos per mass range and the comoving partial volume $\frac{d^2V(z)}{dzd\Omega}$ as follows,
\begin{equation}
    \lambda_{ij} = \Omega_S\int_{z_{j}}^{z_{j+1}} dz \int_{m_{i}}^{m_{i+1}} dm \frac{dn(m,z)}{dm}\frac{d^2V(z)}{dzd\Omega}\,,
    \label{eq:count_prediction}
\end{equation}
where $\Omega_S$ is the survey solid angle. In this work, we use letters of the \textit{latin} alphabet ($i,j,k$...) to refer data vector indices in the data space. Given that the statistics of the underlying matter density field $\delta(\mathbf{x},z)$ are Gaussian, $n(m,z,\mathbf{x})$ is a random Gaussian realisation of $\bar{n}(m, z)$. The observed abundance of clusters $N_{ij}$ in the $i$-th mass and $j$-th redshift bins would be a Poisson realisation of the Gaussian random variable
\begin{equation}
     \widetilde{\lambda}_{ij} = \lambda_{ij}  + \delta\lambda_{ij}\,,
\end{equation}
where the abundance perturbation $\delta\lambda_{ij}$ writes in terms of the halo bias $b$ and the matter overdensity field $\delta$ such as
\begin{align}
    \delta\lambda_{ij} = \int_{\Omega_S} d^2\bm{\theta}\int_{z_{j}}^{z_{j+1}} dz \int_{m_{i}}^{m_{i+1}} & dm \frac{dn(m,z)}{dm}\frac{d^2V(z)}{dzd\Omega} \nonumber\\
    &\times b(m,z) \delta(f_K(w)\bm{\theta},w)\,,
\end{align}
where $\delta(f_K(w)\bm{\theta},w)$ is the three-dimensional matter overdensity field at the comoving position $(x, y, z) = (f_K(w)\bm{\theta},w)$, where $\bm{\theta}$ denotes the two-dimensional position on the sky, and $f_K(w)$ and $w$ are respectively the transverse and line-of-sight comoving distances. The line-of-sight comoving distance depends on the input redshift $w \equiv w(z)$.

For a set of observed cluster abundance $\bm{N} = \{N_{ij}\}$ we have the Gauss-Poisson Compound likelihood likelihood, 

\begin{equation}
\label{eq:mvp_likelihood_multivariate}
    \mathcal{L}(\bm{N}) =\int d\bm{\widetilde{\lambda}} \  \mathcal{N}\qty(\bm{\widetilde{\lambda}}|\bm{\lambda},\Sigma_{\rm SV})\times\prod_{i,j = 1}^c\mathcal{P}\qty(N_{ij} |\widetilde{\lambda}_{ij} )\,,
\end{equation}
where $c = N_z\times N_m$ is the total number of mass-redshift bins. In the above equation,  $\mathcal{N}(\cdot|\bm{\lambda}, \Sigma_{\rm SV})$ is the multivariate Gaussian distribution with mean $\bm{\lambda}= \{\lambda_{ij}\}$ and covariance~$\Sigma_{\rm SV}$. The distribution $\mathcal{P}$ is the single variate Poisson distribution given by
\begin{equation}
\label{eq:poisson_distrib}
\mathcal{P}\qty(N_{ij}|\widetilde{\lambda}_{ij}) = \frac{\widetilde{\lambda}_{ij}^{N_{ij}}}{N_{ij}!}e^{-\widetilde{\lambda}_{ij}}.
\end{equation}
The quantity $\Sigma_{\rm SV}$ is the sample covariance, i.e. the covariance of the $\delta\lambda_{ij}$ such as $\Sigma_{\rm SV}[ij][kl] = \langle \delta\lambda_{ij} \delta\lambda_{kl}\rangle$. 

For a given binning scheme, if we neglect correlations between two different bins of redshift (which in general are small, see \cref{fig:bias_err}), the GPC likelihood in \cref{eq:mvp_likelihood_multivariate} can be written as \citep{2014_Takada}
\begin{equation}
    \mathcal{L}(\bm{N}) = \prod_{j=1}^{N_z} \mathcal{L}_{j}(\{N_{ij}\}_{1\leq i \leq N_m}),
    \label{eq:MPG_block_diag_1}
\end{equation}
where each term $\mathcal{L}_j$ in \cref{eq:MPG_block_diag_1} are given by the integral form
\begin{equation}
\label{eq:MPG_block_diag}
    \mathcal{L}_{j}(\bm{N}) = \int d\delta\ \mathcal{N}(\delta|0,S_{jj})\prod_{i=1}^{N_m} \mathcal{P}(N_{ij}| \lambda_{ij}(1 + b_{ij} \delta))\,.
\end{equation}
In \cref{eq:MPG_block_diag}, $\lambda_{ij}$ is the cluster abundance prediction and $b_{ij}$ is the average halo bias in the $i$-th mass and $j$-th redshift given by $\langle\lambda b \rangle_{ij}/\lambda_{ij}$ in \cref{eq:Nbias}. $S_{jj}=\langle \delta_j^2\rangle$ is fluctuation amplitude of the matter density perturbation $\delta_j$ in the $j$-th redshift bin, as given in \cref{eq:S_ij}. Here, the GPC likelihood is approximated as the product of $N_z$ single variate integrals (instead of computing a $N_z\times N_m$ multivariate integrals).

The GPC satisfies the two limit cases depending on the relative contribution between average cluster abundance shot noise (see \cref{subsec:poisson}) and sample variance $\Sigma_{\rm SV}$, where the GPC likelihood takes the form of a Poissonian distribution or of a Gaussian distribution. The two next sections briefly presents these two limiting cases, which are the most widely used in the literature.

\subsection{Poisson likelihood}
\label{subsec:poisson}
When the cluster abundance is dominant relative to the sample variance, i.e. $\lambda_{ij} \gg \Sigma_{\rm SV}[ij][ij]$ the GPC likelihood in \cref{eq:mvp_likelihood_multivariate} simplifies to the form \citep{Lima_Hu},
\begin{equation}
    \mathcal{L}(\bm{N})\approx \prod_{i,j=1}^{c}\mathcal{P}(N_{ij}|\lambda_{ij})\;,
    \label{eq:poisson_likelihood}
\end{equation}
where $\mathcal{P}$ is the Poisson distribution in \cref{eq:poisson_distrib}, and $\lambda_{ij}$ is the average number of cluster in \cref{eq:count_prediction}. In this approximation, cluster counts in different mass-redshift bins are independent sampling of the Poisson mean $\langle N_{ij}\rangle = \lambda_{ij}$ and each have variance $(\sigma_{ij})^2 =\lambda_{ij}$. This variance $\lambda_{ij}$ is referred to as Poisson shot noise. It denotes the intrinsic scatter of an uncorrelated count experiment.

\subsection{ Gaussian likelihood }
When the shot noise $\lambda_{ij}$ is no more dominant relative to the sample variance, and $\lambda_{ij} \gg 1$, the GPC likelihood takes the form of a multivariate Gaussian likelihood \citep{Lima_Hu}
\begin{equation}
    \mathcal{L}(\bm{N}) \approx \frac{1}{\sqrt{(2\pi)^c}|\Sigma|} \exp \left[ -\frac{1}{2}[\bm{N}-\bm{\lambda}]^T\Sigma^{-1}[\bm{N}-\bm{\lambda}] \right]
    \label{eq:binned_gaussian_likelihood}\,.
\end{equation}
In the above, the full covariance matrix $\Sigma$ for cluster abundance is given by
\begin{equation}
    \Sigma = \Sigma_{\rm SN} + \Sigma_{\rm SV}\,,
    \label{eq:covariance}
\end{equation}
where $\Sigma_{\rm SN} = \mbox{diag}(\bm{\lambda})$ is the Poisson shot noise covariance and is completely diagonal as there is no correlation between different mass-redshift bins. $\Sigma_{\rm SV}$ is the sample covariance. In this approximation, the cluster counts are correlated Gaussian variables with extra-variance given by the sample variance. 

We used the Core Cosmology Library \citep{ChisariCCL2019} to predict the halo mass function from \cite{despali2016universality} and the halo bias from \cite{2010_tinker}.


\section{Testing likelihood accuracy}
\label{section:likelihood_accuracy}

To take an explicit example, we can consider multiple realisations of our Universe which have at their origin the exact same cosmological model. For each realisation, we use a cluster abundance likelihood to estimate a set of cosmological parameters. In the case that we use the {\it true} model describing the statistics of cluster abundance in the Universe, the posteriors we estimate in each Universe will be consistent, after accounting for the expected statistical fluctuations. If our cluster abundance likelihood is lacking, for example, the Poisson likelihood missing the sample variance contribution, our estimated parameters may be biased and/or misestimate the parameter errors. In this case, the analysis of two realisations of the same cosmological model may lead to the estimation of inconsistent cosmological parameters. 

Therefore we can develop a test for the likelihood model by comparing the dispersion of the posterior means over  many realisations of the Universe against the predicted dispersion from an individual realisation. And more precisely, we can compare the average posterior covariance from an ensemble of the simulations to the covariance of an ensemble of posterior means. 

The posterior distribution for a set of parameters $\bm{\theta} = \{\theta_\alpha\}$ is given in terms of the likelihood $\mathcal{L}_Y$ and the prior $\pi(\bm{\theta})$ by the Bayes theorem, namely
\begin{equation}
    \mathcal{P}_Y(\bm{\theta}|\bm{N}) \propto \mathcal{L}_Y(\bm{N}|\bm{\theta})\pi(\bm{\theta}).
    \label{eq:Posterior}
\end{equation}
where $\mathcal{L}_Y$ is the likelihood used for cosmological inference, i.e. the one \textit{chosen} for the analysis. For an \textit{individual} dataset $\bm{N}$, the posterior covariance for a set of parameters $\bm{\theta}$ estimated from $\mathcal{P}_Y$ is defined as
\begin{equation}
    \mathcal{C}^{\rm{ind}}_{\alpha \beta} = \mathbb{E}_{\mathcal{P}_Y} 
    [ ( \theta_\alpha - \widehat{\theta}_\alpha ) ( \theta_\beta - \widehat{\theta}_\beta ) ],
    \label{eq:individual_covariance}
\end{equation}
where $\widehat{\theta}_\alpha$ is the average parameter given the posterior $\mathcal{P}_Y$ namely 
\begin{equation}
     \widehat{\theta}_\alpha = \mathbb{E}_{\mathcal{P}_Y}[ \theta_\alpha ].
     \label{eq:estimator_mean}
\end{equation}
In this work, we will consider \textit{greek} symbols ($\alpha, \beta, \gamma$, ...) to refer to parameter indices in the cosmological parameter space.
A lower bound for the posterior covariance, $\mathcal{C}^{\rm{ind}}$, can be set using the Fisher matrix formalism \cite{FISHER} (see also \citealt{tegmark} and \citealt{heavens_2014}) and the Cramer-Rao inequality \citep{CRRau}
\begin{equation}
    \mathcal{C}^{\rm{ind}}_{\alpha\beta} \geq (\mathrm{F}^{-1})_{\alpha\beta},
\end{equation} 
where $\mathrm{F}$ is the Fisher matrix given by \citep{FISHER},
\begin{equation}
    \mathrm{F}_{\alpha\beta} = \mathbb{E}_{\mathcal{L}_Y}
     \left[ \left( \frac{\partial}{\partial\theta_{\alpha}} \ln \mathcal{L}_Y \right)
     \left( \frac{\partial}{\partial\theta_{\beta}} \ln \mathcal{L}_Y \right) \right],
    \label{eq:Fisher_formula}
\end{equation}
where the expectation runs over the realisation of the data that follows the likelihood distribution. The Fisher information matrix for multivariate Gaussian and Poisson likelihoods (see \cref{app:var_est_2}) is,
\begin{equation}
    \mathrm{F}_{\alpha\beta} = \lambda_{,\alpha}^T\Sigma_Y^{-1} \lambda_{,\beta}\;,
    \label{eq:Fisher_formula_gaussian}
\end{equation}
with $\Sigma_Y$ the data covariance matrix assumed to be independent of the cosmological parameters, and where $\lambda_{,\alpha}$ is the derivative of the model that predict the signal with respect to the cosmological parameter $\alpha$. The Fisher forecast estimates the first order of the posterior covariance. When the posterior is not Gaussian in the parameter space, the first order Fisher forecast is not sufficient and higher order forecasts needs to be considered \citep{Wolz_2012,Sellentinforecast14}. In our case, we will see that posteriors are almost Gaussian in the parameter space, then $\mathrm{F}_{\alpha\beta}^{-1}\approx  \mathcal{C}^{\rm{ind}}_{\alpha\beta}$.

We now define $\mathcal{C}^{\rm{ens}}$ as the covariance of the ensemble of posterior means obtained for data generated from the latent likelihood $\mathcal{L}_X$. Therefore,

\begin{equation}
\mathcal{C}^{\rm{ens}}_{\alpha \beta} = \mathbb{E}_{\mathcal{L}_X} \left[ \left( \left< \widehat{\theta}_\alpha \right> -\widehat{\theta}_\alpha \right) \left ( \left< \widehat{\theta}_\beta \right> -\widehat{\theta}_\beta \right) \right]\,,
 \label{eq:cov_ensemble}
\end{equation}
where $\widehat{\theta}_\alpha$ is the individual posterior mean defined in \mbox{\cref{eq:estimator_mean}}, derived from the individual posterior $\mathcal{P}_Y$ in \cref{eq:Posterior}. We have introduced the likelihood $\mathcal{L}_{X}$, which is the true underlying likelihood which describes the data. By definition, $\mathcal{C}^{\rm{ens}}$ defines the frequentist covariance of a Bayesian estimator, that depends on both analysis and underlying likelihoods $\mathcal{L}_Y$ and $\mathcal{L}_X$ (respectively from $\widehat{\theta}$ in \cref{eq:estimator_mean} and the average $\mathbb{E}_{\mathcal{L}_X}[\cdot]$). We can then show that, if $\Sigma_X$ denotes the data covariance matrix derived from the latent distribution $\mathcal{L}_X$ (for example in the case of a Gaussian latent likelihood), the ensemble covariance $\mathcal{C}^{\rm ens}$ can be forecasted (see \cref{app:var_est} for details) such as 
\begin{equation}
 \mathcal{C}_{\alpha\beta}^{\mathrm{ens}} = (\mathcal{C}^{\mathrm{ind}} \lambda_{,})_\alpha^T\Sigma_Y^{-1}\Sigma_X\Sigma_Y^{-1}(\mathcal{C}^{\mathrm{ind}} \lambda_{,})_\beta\,.
 \label{eq:cov_real_formula}
\end{equation}
Finally, if we substitute \cref{eq:Fisher_formula_gaussian} into \cref{eq:cov_real_formula} and take the two covariance matrices to be equal\footnote{This is the case when the latent likelihood and the likelihood used for the posterior estimation parameters are the same.}, $\Sigma_Y = \Sigma_X$ (see again \cref{app:var_est}), then the two posteriors are indeed equivalent,
\begin{equation}
    \mathcal{C}^{\mathrm{ind}}=\mathcal{C}^{\mathrm{ens}}\;.
\end{equation}
We can therefore justify the comparison of these two different posterior covariance matrices as a method to test the accuracy of a given likelihood. Additionally in the case that $\Sigma_Y \neq \Sigma_X$ we have,
\begin{equation}
    \mathcal{C}^{\mathrm{ind}} \neq \mathcal{C}^{\mathrm{ens}}.
\end{equation}

In the following we assume this result also holds for the Gauss-Poisson likelihood (although not formally demonstrated). Since we find the latent likelihood to be close to Gaussian this is likely true in our specific case. 

Note that this equivalence is discussed more general terms in \cite{Efron}, and also in \cite{Penna_Lima_2014}
for an unbinned Poisson likelihood. Similar calculations are made in \cite{percival2022matching} in order to obtain a prior which gives equivalent results for Frequentist and Bayesian analyses.

In order to perform such tests we require many simulations of the universe. These simulations are described in the following section.

\section{The PINOCCHIO catalogs}
\label{section:pinocchio}

Full N-body numerical simulations are computationally expensive. Therefore, in order to have many simulated realisations of the Universe to test the cluster abundance likelihood, we use a set of halo catalogs generated by an approximate method called PINOCCHIO (PINpointing Orbit-Crossing Collapsed HIerarchical Objects) \citep{monaco2002pinocchio,munari2017improving}. PINOCCHIO is an algorithm which generates dark matter halo catalogs using Lagrangian Perturbation Theory \citep{moutarde1991precollapse,buchert1992lagrangian,bouchet1994perturbative} and ellipsoidal collapse (\citet{bond1996peak,eisenstein1994analytical}, \citet{Monaco1997}). Relative to N-body simulations, the accuracy of the PINOCCHIO algorithm is within $\sim 5-10 \%$ for reproducing the halo mass function, halo two-point statistics, and the halo bias at all relevant redshifts used in this study \citep{munari2017improving}.

We use a set of 1000 light-cone catalogs which cover a quarter of the sky, a redshift range of $z=0$ to $z=2.5$ and halos with virial masses above $2.45 \times 10^{13}$\;$h^{-1}$M$_\odot$. The simulated cosmology is the best-fit cosmology from \citet{Planck_14}: $\Omega_m$ = 0.30711, $\Omega_b$ = 0.048254, h = 0.6777, $n_s$ = 0.96, $\sigma_8$= 0.8288.  These same catalogs were used in the analysis of \cite{Fumagalli_2021}. To bypass the small inaccuracy of the PINOCCHIO halo mass function, PINOCCHIO halo masses have been rescaled so as to abundance-match the mass function averaged over 1000 realisations with the \citet{despali2016universality} analytic fit. This procedure allows us to preserve sample variance of the individual realisations while forcing the average to reproduce a known function.

\begin{figure*}
    \centering
    \includegraphics[width=.8\textwidth]{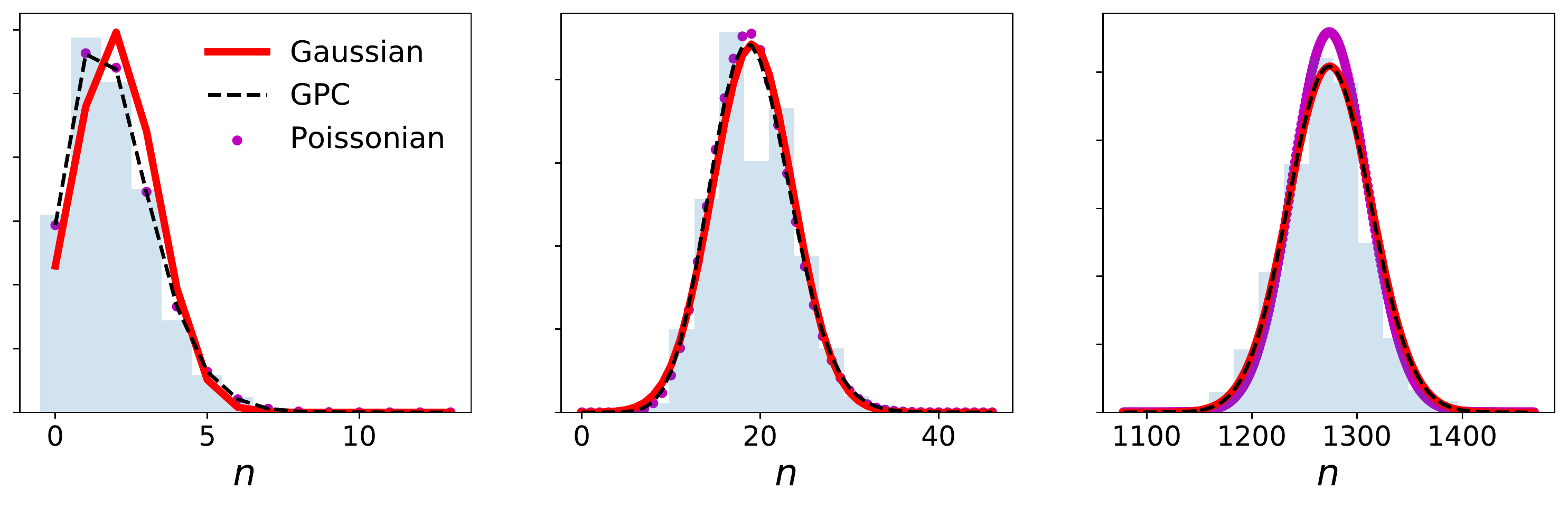}
    \caption{Blue histograms: distribution of the cluster counts over the 1000 PINOCCHIO simulations for a low- (left), intermediate- (middle) and highly-populated bin (right). Magenta symbols: Single-variate Poisson distribution from \cref{eq:poisson_likelihood}. Red line: single variate Gaussian distribution from \cref{eq:binned_gaussian_likelihood}. Black dashed line: GPC likelihood from \cref{eq:mvp_likelihood_multivariate}.  }
    \label{fig:pinocchio_dstrib}
\end{figure*}

In \cref{fig:pinocchio_dstrib}, we consider 3 different mass-redshift bins, corresponding respectively to a low, medium, high average number of dark matter halos (from left to right $\langle N\rangle\sim 2$, $\langle N\rangle\sim 20$ and $\langle N\rangle\sim 1280$). For each of these bins, we make the histogram of cluster count over the 1000 PINOCCHIO dark matter catalogs (shaded regions). Doing so we have estimates of the cluster count likelihoods in each of those three bins. For each binning, we plot the 3 different likelihoods discussed in \cref{section:cluster_distributions},  the Poissonian (purple dots), the Gaussian (red line) and the Gauss-Poisson Compound (black dashed line). We see for the small population bins that the distribution of the cluster abundance in small bins is non-Gaussian; additionally we see for large population bins that the distribution is non-Poissonian due to the sample variance contribution. 

\section{Likelihood covariance for the Gaussian and GPC cases}
\label{sec:covariance}

For the Gaussian and GPC likelihoods, the count covariance $\Sigma$ needs to be calculated. The abundance covariance can be predicted from theoretical considerations and some simplifying approximations, for a given cosmology, mass function and halo bias prescriptions. As mentioned in \cref{section:likelihood_accuracy}, the count covariance has two contributions $\Sigma^{\rm th} = \Sigma_{\rm SN}^{\rm th} + \Sigma_{\rm SV}^{\rm th}$, where $\Sigma_{\rm SN}^{\rm th}$ is the shot noise variance and $\Sigma_{\rm SV}^{\rm th}$ is the sample covariance. 

The shot noise contribution in the $i$-th mass bin and $j$-th redshift bin is equal to the estimated count $\lambda_{ij}$ in the considered mass-redshift bin, as given by equation \cref{eq:count_prediction}, i.e.
\begin{equation}
    \Sigma_{\rm SN}^{\rm th}[ij][kl] = \delta^K_{ik}\delta^K_{jl} \lambda_{ij}\;,
    \label{eq:sn_theo}
\end{equation}
where $\delta^K_{ij}$ is the Kronecker Delta function. As such, it depends on the cosmology and on the halo mass function.

The sample covariance $\Sigma_{\rm SV}$ between 2 counts $N_{ij}$ and $N_{kl}$, in two mass-redshift bins, is computed using \citep{2018_Lacasa_SSC_partial_sky}
\begin{equation}
   \Sigma_{\rm SV}^{\rm th}[ij][kl] \approx \langle b\lambda\rangle_{ij} \langle b\lambda\rangle_{kl} S_{jl}\;,
\label{eq:sv_theo}
\end{equation}
where $S_{jl}$ is the matter density fluctuation covariance between the two redshift bins $j$ and $l$. In practice, we use the \texttt{PySSC} package\footnote{\url{ https://github.com/fabienlacasa/PySSC}} to compute $S_{jl}$ (see \cref{app:sample_variance_pyssc} for details). The quantity $\langle b\lambda\rangle_{ij}$ is the average halo bias in the $i$-th mass and $j$-th redshift bin such as
\begin{equation}
    \langle b\lambda\rangle_{ij}=\Omega_S\int_{z_j}^{z_{j+1}} dz \int_{m_{i}}^{m_{i+1}} dm \frac{dn(m,z)}{dm}\frac{d^2V(z)}{dzd\Omega}b(m,z)';.
    \label{eq:Nbias}
\end{equation}

The cluster count covariance matrix may also be estimated from the data. To validate our use of the \texttt{PySSC} code, Appendix \ref{app:sample_variance_pyssc} presents the comparison between the covariance matrix estimated from the 1000 simulations to the theoretical prediction computed with \texttt{PySSC}.

\section{Methodology}
\label{sec:methodology}

The method we use to test each likelihood is the following:

\begin{itemize}
    \item for each of the 1000 mock catalogs we obtain a posterior (using importance sampling); 
    \item from each of these posteriors we calculate the posterior mean and posterior covariance;
    \item from the ensemble of the posteriors we calculate the mean of the posterior means and the covariance of the posterior means;
    \item we then compare the results of the individual posteriors against the ensemble of the posteriors;
    \item we compute the Fisher forecast from \cref{eq:Fisher_formula_gaussian} and ensemble forecasts from \cref{eq:cov_real_formula} for comparison.
\end{itemize} 

This test is performed for each of the likelihoods for three different binning schemes. The binning schemes in the mass-redshift plane are given in \cref{subsec:binning}. The method used, importance sampling, is detailed in \cref{subsec:IS}. In \cref{subsec:posterior_robustness} we show how the ensemble posterior covariance is calculated.

\subsection{Mass and redshift ranges, and binning setups}
\label{subsec:binning}
\begin{figure*}
    \centering
    \includegraphics[width=.9\textwidth]{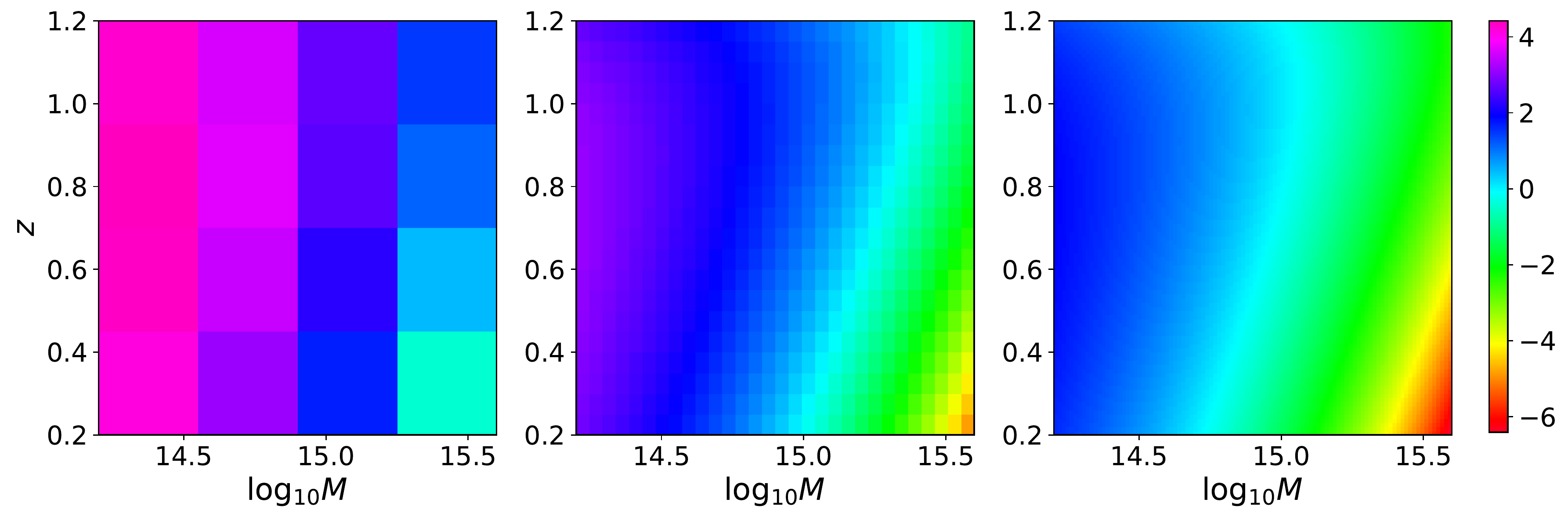}
    \caption{2D histograms of $\log_{10}\lambda_{ij}$ abundance predictions ($i$-th mass and $j$-th redshift bin) for the {\tt coarse} (left), {\tt medium} (center), and {\tt fine} (right) binning schemes of \cref{tab:binnings}.}
    \label{fig:binning_scheme}
\end{figure*}

For this analysis, we adopt a binning scheme of the $(\log_{10})$mass-redshift plane within the redshift range $[z_{\rm min}, z_{\rm max}]= [0.2, 1.2]$ and the mass range $[\log_{10} (m_{\rm min}/M_\odot), \log_{10}(m_{\rm max}/M_\odot)]= [14.2, 15.6]$. We have three different binnings; a { \tt coarse}, { \tt medium} and {\tt fine} binning which are detailed in \cref{tab:binnings}. Both redshift and $(\log_{10})$mass bins are equally spaced.

\begin{table}
\begin{center}
\begin{tabular}{l|c|c|c } 
\hline
 & \tt{coarse} & \tt{medium} & \tt{fine} \\ 
\hline\hline
Number of redshift bins $N_{z}$ & 4 & 20 & 100 \\ 
Number of mass bins $N_{m}$ & 4 & 30 & 100 \\ 
Total number of bins $N_{\rm tot}$ & 16 & 600 & 10000\\
\hline
\end{tabular}
\caption{Characteristics of the {\tt coarse}, {\tt medium} and {\tt fine} two-dimensional binnings of the mass-redshift plane chosen in this analysis.}
\label{tab:binnings}
\end{center}
\end{table}

\Cref{fig:binning_scheme} shows the predicted abundance for the fiducial cosmology for the three binning schemes. For the  {\tt coarse} case, the number of expected halos per bin never drops below one and the bins are generally well-populated (there is an average of $5000$ halos per mass and redshift over all the binning scheme, ranging from 1 to 25 000). For the {\tt medium} and {\tt fine} grids a significant portion of the mass-redshift plane are expected to have only a few or no halos. From this we expect the Gaussian approximation to be valid only for the {\tt coarse} binning scheme, while this may be a less accurate statistical description for the two other cases. The different binning schemes allow us to explore a variety of regimes, from the shot noise-dominated regime to the non-negligible sample variance regime, and to test the performance of the likelihoods described in \cref{section:cluster_distributions} in each regime.

\subsection{Posterior estimation using importance sampling}
\label{subsec:IS}

To undertake this analysis, from the cluster abundance data vector $\widehat N$, we perform a cosmological analysis for the 1000 simulations, three binning schemes and three likelihoods. Therefore we wish to obtain 9000 posteriors in the parameter space $\boldsymbol{\theta}=(\Omega_m, \sigma_8)$. This would be computationally expensive for a standard Markov Chain Monte Carlo method. Instead, we use importance sampling (IS) \footnote{Although not shown here, we did check on a single simulation that the two approaches yield the same results. Importance sampling is however much faster to run, since count prediction is evaluated once at each location in the parameter space. In particular, it is much faster for GPC where the likelihood function is computationally taxing.} to efficiently sample the joint posterior distribution of the parameters. We briefly recap the basic framework of importance sampling below (see \citealt{2021arXiv210205407E} for a recent review).

Importance sampling can be used to estimate the statistical properties of the posterior distribution $\mathcal{P}_Y(\bm{\theta}|\bm{N})$ in \cref{eq:Posterior} from random sample of points that follows a proposal distribution $q$. Considering a random variable $X \sim p$, the expectation of $X$ is given by

\begin{equation}
    \mathrm{E}_p[X] = \int_{-\infty}^{+\infty} dx p(x) x.
\end{equation}
The above average can be estimated from sampling another random variable $Y$ following a proposal distribution $q$. We call $\widehat{X}$ the unbiased estimator of $\mathrm{E}_p[X]$,

\begin{equation}
    \widehat{X}=\frac{1}{N_q}\sum_{k=1}^{N_q} w_k y_k,
\end{equation}
where $N_q$ is the number of samples and the weights, $w_k$ are defined as,
\begin{equation}
    w_k = \frac{p(y_k)}{q(y_k)}.
\end{equation}
The mean of $X$ is therefore expressed as a weighted mean over the sample $Y$. So the estimated individual average parameter $\widehat{\theta}^{\rm ind}_\alpha$ is given by,

\begin{equation}
    \widehat{\theta}^{\rm ind}_\alpha=\sum_{k=1}^{N_q}w_k[\theta_\alpha]_k,
    \label{eq:best_fit_IS}
\end{equation}
where $[\theta_\alpha]_k$ is the sampled parameter value $\theta_\alpha$ at position $k$.
From this, one may also estimate the unbiased weighted covariance matrix of $ \boldsymbol{\theta} = (\Omega_m, \sigma_8)$ as

\begin{equation}
\widehat{\mathrm{C}^{\rm ind}_{\alpha\beta}} = \sum_{k=1}^{N_q}\widetilde{w}_k([\theta_\alpha]_k-\widehat{\theta}^{\rm ind}_\alpha)([\theta_\beta]_k-\widehat{\theta}^{\rm ind}_\beta).
\label{eq:cov_IS}
\end{equation}

with

\begin{equation}
    \widetilde{w}_k= w_k\frac{\sum_{k=1}^{N_q}w_k}{\left(\sum_{k=1}^{N_q}w_k\right)^2-\sum_{k=1}^{N_q}w_k^2}.
\end{equation}

For the proposal $q$, we use a bi-variate Gaussian distribution centered on the $(\Omega_m, \sigma_8)$ input values of the PINOCCHIO simulations. The parameter covariance matrix of the proposal distribution $q$ is chosen as the Fisher forecast (given in \cref{eq:Fisher_formula_gaussian}) for the $\tt{coarse}$ binning scheme, multiplied by a factor 36. This ensures that the proposal distribution encompasses a large enough region of the parameter space covered by the ensemble of $\sigma$ contours of each individual posterior. 

To compute the posterior we use the different likelihoods $\mathcal{L}$ presented in \cref{section:cluster_distributions}, and we use flat priors for the parameters, namely $\Omega_m \in [0.293, 0.321]$ and $\sigma_8\in[0.820, 0.836]$. These ranges were chosen as $12$ times the square root of the Fisher variances on each parameter given in \cref{eq:Fisher_formula_gaussian}. The proposal has to be \textit{wide} enough to contain $2\sigma$ contours for each posterior. 

\begin{figure}
    \centering
    \includegraphics[width=0.9\columnwidth]{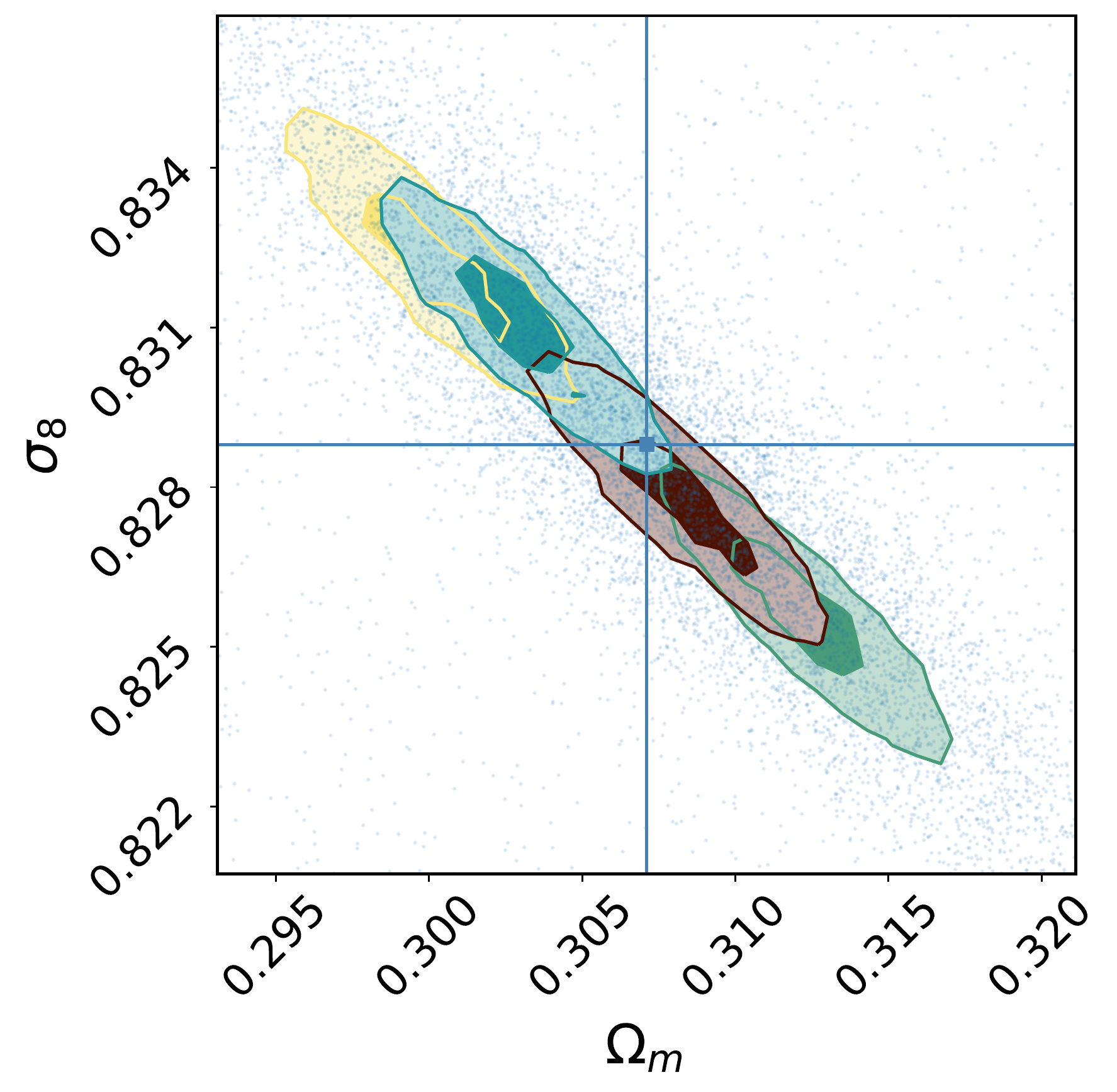}
    \caption{Illustration of the importance sampling input and output. Background blue points: random sample of $10^4$ points following the proposal distribution $q$. The contour plots correspond to the 1 and 2$-\sigma$ levels of the posterior distributions of $\Omega_m$ and $\sigma_8$, for four different PINOCCHIO simulations. Vertical and horizontal lines correspond to the input parameters used in the PINOCCHIO simulations.}
    \label{fig:importance_sampling}
\end{figure}

As an illustration, the light blue points in \cref{fig:importance_sampling} show $N_q=10^4$ samples of the proposal distribution $q$ for the { \tt coarse} $4\times 4$ binning. The four coloured contours corresponds to the 1 and 2$\sigma$ confidence intervals obtained using the binned Poissonian likelihood, i.e. ${\cal L}(N|\theta)$ given by \cref{eq:poisson_likelihood}, for four of the PINOCCHIO simulations\footnote{The codes used to compute the cluster abundance and covariance predictions, the likelihood, the forecasts and run the importance sampling are available at \url{https://github.com/payerne/LikelihoodsClusterAbundance.}}. These are obtained as the weighted histograms of the $q$-sample. The unsmoothed contours have been obtained directly from the $10^4$ samples. 

\subsection{Estimating the ensemble posterior covariance}
\label{subsec:posterior_robustness}

After the 1000 posteriors have been estimated using importance sampling for each of the 1000 simulations we can estimate the ensemble posterior covariance. The estimator of the cosmological parameter mean value $\widehat\theta_{\alpha}$ in \cref{eq:estimator_mean} is noted $[\widehat{\theta}^{\rm ind}_\alpha]_k$ for the $k$-th simulation and is calculated as in \cref{eq:best_fit_IS}. The mean cosmological parameters over the ensemble $\langle \widehat{\theta}_\alpha\rangle$ that appear in \cref{eq:cov_ensemble} are calculated from these mean values,
\begin{equation}
      \widehat{\theta}^{\rm ens}_\alpha=\frac{1}{N_{\rm sim}}\sum_{k=1}^{N_{\rm sim}}[\widehat{\theta}^{\rm ind}_\alpha]_k.
\end{equation}
The corresponding ensemble covariance matrix is estimated as,
  \begin{equation}
      \widehat{\mathcal{C}^{\rm ens}_{\alpha \beta}} = \frac{1}{N_{\rm sim}-1}\sum_{k=1}^{N_{\rm sim}}([\widehat{\theta}^{\rm ind}_\alpha]_k - \widehat{\theta}^{\rm ens}_\alpha)([\widehat{\theta}^{\rm ind}_\beta]_k - \widehat{\theta}^{\rm ens}_\beta)\;.
      \label{eq:C_param_ens}
  \end{equation}
From this covariance matrix we can extract the ensemble variance and correlation coefficient in order to compare these results against those of the individual posteriors calculated in \cref{eq:cov_IS}.

\section{Results}
\label{section:results}

In this Section we present the results of testing the accuracy of likelihoods using importance sampling and the set of the 1000 PINOCCHIO simulations. In \cref{subsec:bestfit} we compare the individual posterior means to the ensemble posterior mean, in order to check for a bias in the likelihoods. In \cref{subsec:cov_robustness_measure} we compare the individual posterior variances and individual posterior correlation coefficients against the ensemble posterior variance and ensemble correlation coefficients, in order to show the robustness of the estimated posteriors. In \cref{subsec:partial} and \cref{subsec:highmass} we respectively reduce the sky area and increase the lower mass-cut value of the analysis in order to identify if a regime exists in which the Poisson likelihood or the GPC likelihood obtain tighter, yet still robust, cosmological constraints than the Gaussian likelihood.

\subsection{Bias on the posterior mean}
\label{subsec:bestfit}

\begin{figure}
    \includegraphics[width=\columnwidth]{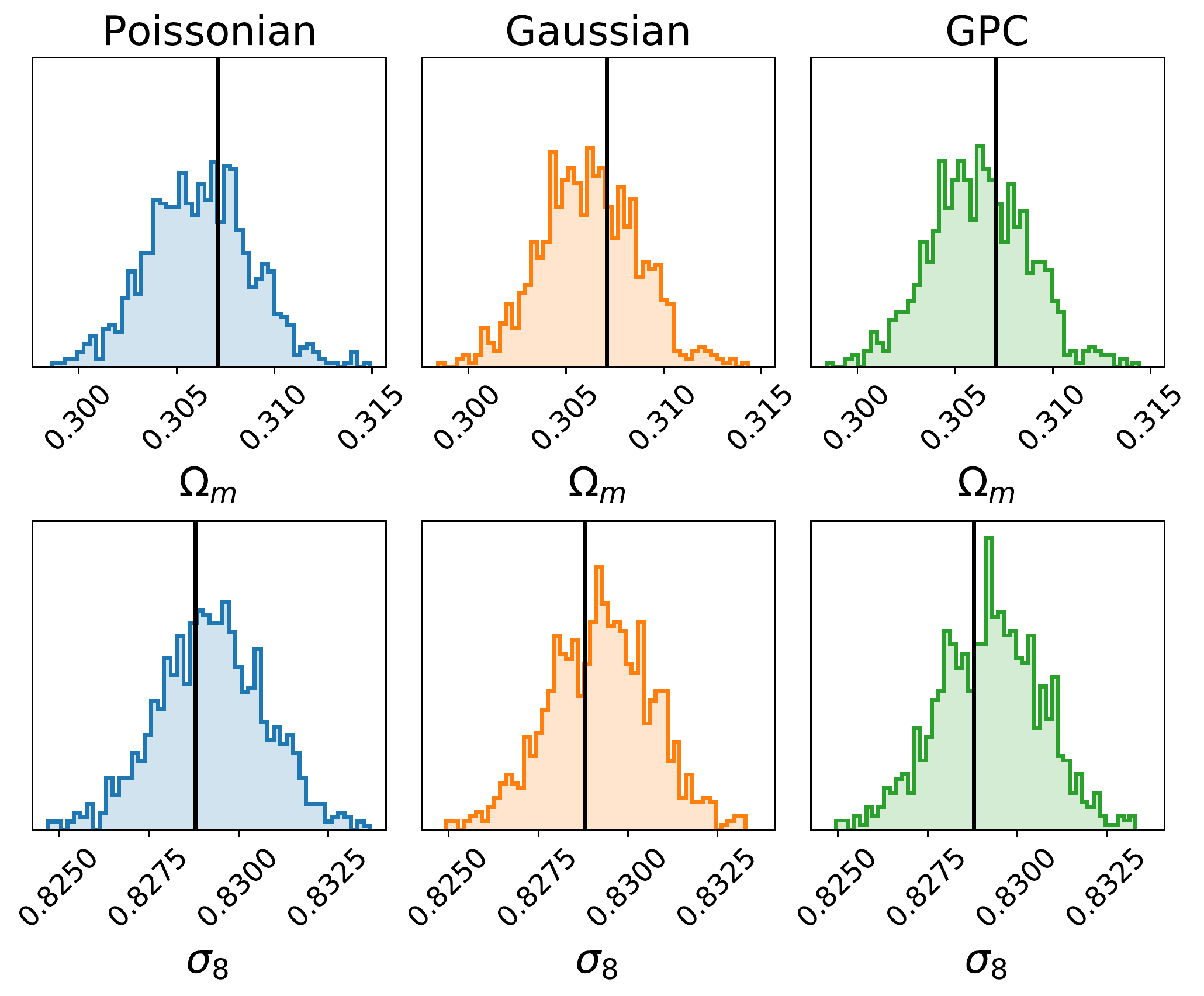}
    \caption{Histogram of the 1000 posterior means of $\Omega_m$ (top) and $\sigma_8$ (bottom) for the 3 likelihoods (columns) for the first binning setup in \cref{tab:binnings}. The vertical black lines correspond to the input cosmology of the PINOCCHIO simulations.}
    \label{fig:4z4m_bfit_full}
\end{figure}

\Cref{fig:4z4m_bfit_full} shows, for the {\tt coarse} $4\times 4$ binning, the histograms of the 1000 means of the posteriors for $\Omega_m$ (top row) and $\sigma_8$ (bottom row), computed with \cref{eq:best_fit_IS}. From left to right, the columns corresponds to the three likelihoods; Poissonian, Gaussian and GPC. The input values of the parameters are given by the vertical solid lines. The distributions of the estimated cosmological parameters are very similar for each likelihood and they are roughly normally distributed about the input cosmology.

\begin{figure}
    \includegraphics[width=\columnwidth]{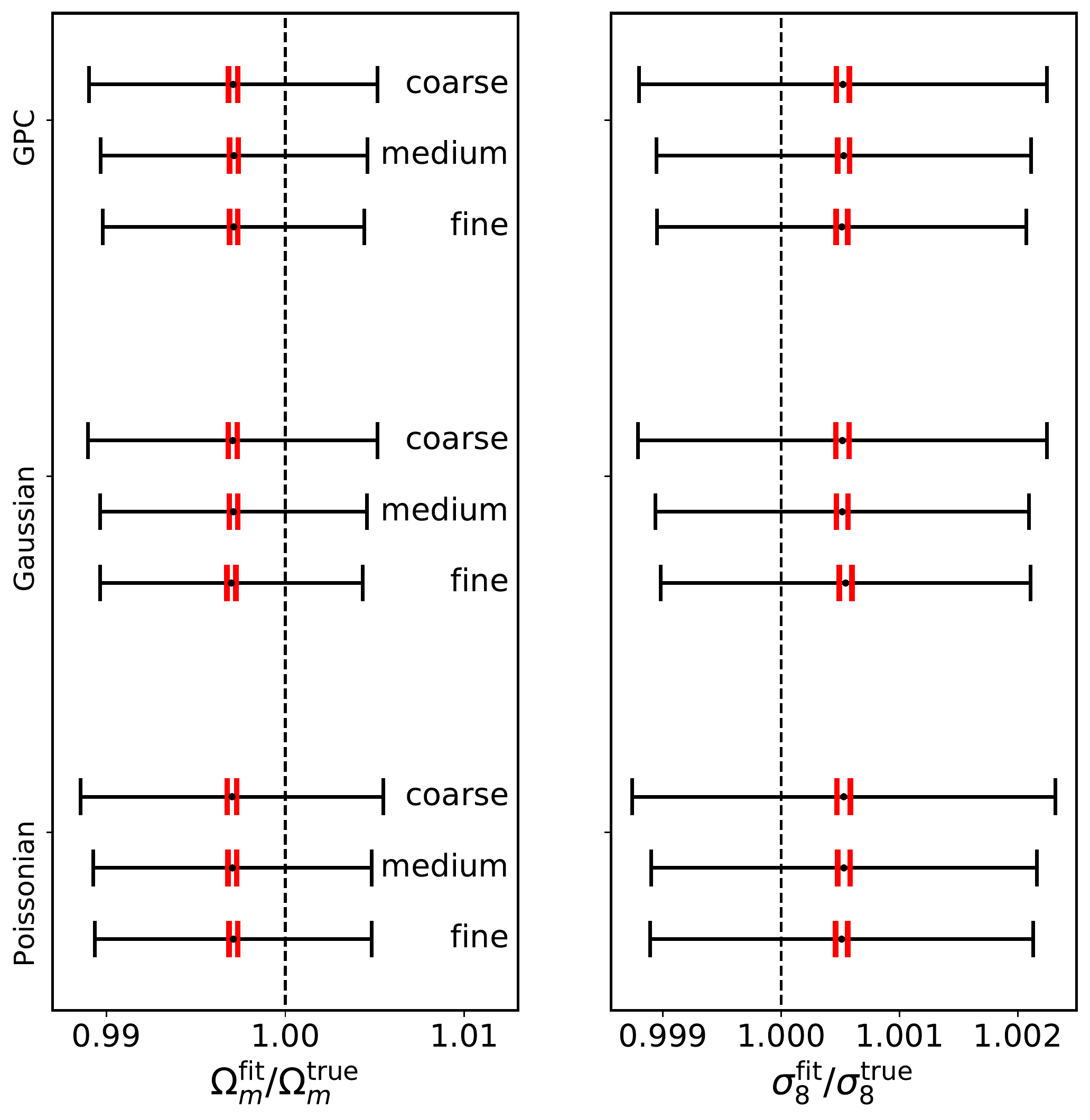}
    \caption{Average of individual posterior means for $\Omega_m$ and $\sigma_8$ (left and right respectively) with standard deviations of posterior means (black error bars) and errors on the mean (red error bars) for the three likelihoods and the three binning setups in \cref{tab:binnings}.}
    \label{fig:bfit_summary}
\end{figure}

In \cref{fig:bfit_summary} we summarise the results for the bias on the mean estimated cosmological parameters for each of the three binnings and the three likelihoods. The black dots show the mean of the 1000 estimated cosmological parameters. The black solid lines shows the standard deviation of the mean. The red intervals shows the error on the mean of the 1000 estimated cosmological parameters (the standard deviation divided by $\sqrt{1000}$ ). The black dotted line indicates the input cosmological parameters to the simulations. We see that there is a small consistent bias between the input value of the cosmological parameters and the mean of the estimated parameters (the dotted black line does not fall within the red error bars for any of the binnings or the likelihoods). There are many potential explanations for this bias. It suggests that the likelihood used and therefore the underlying halo model is on some level inaccurate. As the bias is consistent between the Poisson likelihood which does not use the halo bias and the other two likelihoods which do this bias is not because of the halo bias model used. The bias is nonetheless small, as it is sub-percent and much smaller than the variance of an individual posterior, which we present in the following subsection.

\subsection{Robustness of the posterior covariance}
\label{subsec:cov_robustness_measure}

Now we consider the errors and correlations of the parameters obtained from individual simulations by comparing them to the dispersion of the mean values of the posteriors over the 1000 simulations. As discussed in \cref{section:likelihood_accuracy}, this provides a useful test to check the robustness of the error bars. 

\begin{figure*}
    \centering
    \includegraphics[width=.8\textwidth]{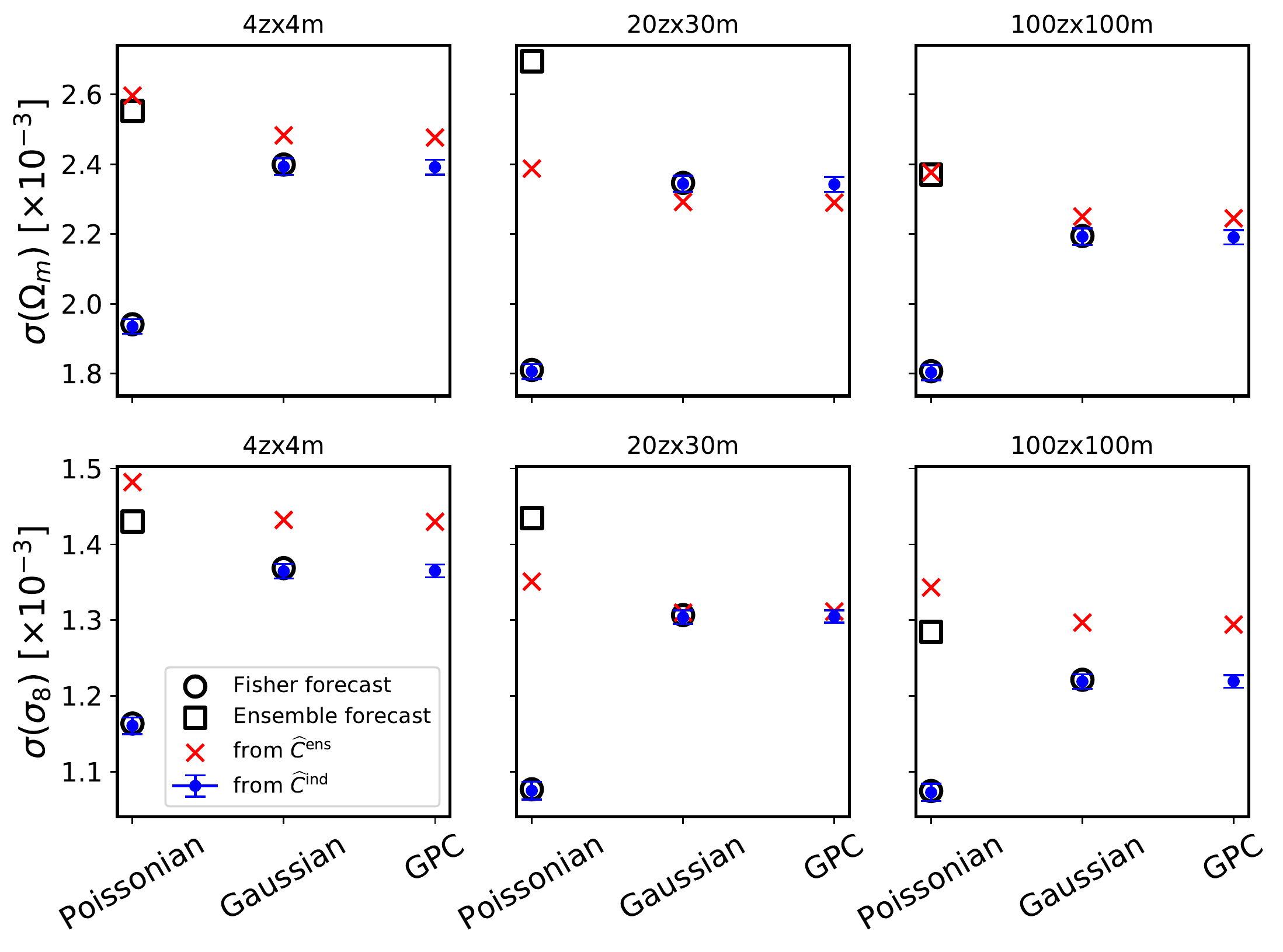}
    \caption{Errors on $\Omega_m$ (top) and $\sigma_8$ (bottom) for the different binning setups in \cref{tab:binnings} (columns). For each plots, the chosen likelihood is given on the x-axis while errors are on the y-axis. Blue dots correspond to the average of individual errors, and the error-bars represent the standard deviation of these individual errors. Red crosses correspond to the standard deviation of the 1000 best fits. Black circles correspond to Fisher forecasts in \cref{eq:Fisher_formula_gaussian} (for the Poissonian and Gaussian likelihoods) and black boxes to Ensemble forecast in \cref{eq:cov_real_formula} (only for the Poisson likelihood, see text for details).}
    \label{fig:std_full}
    \centering
\end{figure*}

\Cref{fig:std_full} summarizes the error estimations for the three binning schemes (columns) and the two parameters ($\sigma_8$ and $\Omega_m$ on the top and bottom row respectively). In each panel, the likelihood used for the analysis is given on the x-axis. The blue dots with errorbars correspond to the mean and dispersion of the individual errors that were computed using the importance sampling covariance matrix in \cref{eq:cov_IS}. The red crosses are the values of the standard deviation of the ensemble of mean values, calculated from the ensemble covariance matrix presented in \cref{eq:C_param_ens}. The red crosses are the same errors as the black error bars in \cref{fig:bfit_summary}.

The errors slightly decrease when using more bins, regardless of the likelihood, with an improvement of ~10\%, between the {\tt coarse} (left) and {\tt medium} (middle) binning scheme. There might therefore be a small benefit to consider a finer binning when dealing with the upcoming survey data of Rubin LSST or {\it Euclid}. Pushing things further, there seems however to be no gain (or only a very marginal one) in using the {\tt fine} binning compared to the {\tt medium} one. 

The Poisson likelihood underestimates the individual errors by up to $30\%$. This is evident as the blue dots are below the red crosses (the individual errors are smaller than the ensemble errors). This is true even when a significant fraction of the mass-redshift space has low number counts for {\tt medium} and {\tt fine} binnings.  See for example \cref{fig:binning_scheme}. The errors are underestimated for the Poisson likelihood as this likelihood does not include the sample variance contribution to the cluster abundance. 

We see that the agreement is much better for the Gaussian and GPC descriptions, that account for sample variance. The individual errors are only slightly under- or over-estimated ($\sim 2-5\%$ for the {\tt coarse} and {\tt medium} cases) compared to the dispersion of the 1000 means. It seems likely that this slight underestimation arises from our use of only the block-diagonal terms in the Gaussian and GPC covariance matrices (we ignore correlations between redshift bins).

In \cite{Fumagalli_2021} it is also evident that the parameter constraints for the average individual posteriors for the Poisson likelihood are tighter than that of the Gaussian likelihood. In this work we were able to show explicitly that this is not because the Poisson uses information beyond the Gaussian likelihood but rather that the Poisson neglects the sample variance contribution.

We can also better understand the results using two different forecasts, motivated by the discussion presented in \cref{section:likelihood_accuracy}. We compare the individual posterior results and the ensemble posterior results against two different forecasts. The first is the often used Fisher forecast \cite[e.g., ][]{sartoris, artis_2021}, introduced here in \cref{eq:Fisher_formula_gaussian}. This forecast should always agree with the estimated individual errors since the likelihood used in the analysis (which we call the {\it analysis} likelihood) is assumed to be equivalent to the latent likelihood driving the data. These forecasts are the black circles in \cref{fig:std_full}. There is agreement between the blue dots and the black circles therefore these forecasts provide a consistency check on the analysis. We did not compute the Fisher forecast of the GPC individual parameter covariance, since it requires the computation of as many multidimensional integrals as the total number of mass-redshift bins in \cref{eq:Fisher_formula}, so we did not investigate this complex computation in this work. But regarding the difference between Gaussian and GPC individual errors for all binning scheme, we can assume that the GPC Fisher constraints are roughly the same as for the Gaussian case.

Secondly we look at the ensemble forecast introduced in \cref{eq:cov_real_formula}. This gives a prediction of the errors on the cosmological parameters where the analysis likelihood differs from the latent likelihood. As we do not know the latent likelihood we use the Gaussian likelihood as a substitute for the latent likelihood in this forecast, therefore stating explicitly that $\mathcal{C}^{\rm ens} = \mathcal{C}^{\rm ind}$ for the Gaussian analysis likelihood. These forecast are shown for the Poisson analysis likelihood with the black boxes in \cref{fig:std_full}.  We found that for the Poisson analysis likelihood these forecasts predict that the ensemble variance is higher than individual errors. Still for the Poisson likelihood, both ensemble and Fisher forecast are respectively larger and smaller than the \textit{correct} errors we should obtain if the Gaussian likelihood is the latent one (black circle in the Gaussian column). We do not show the ensemble forecast for the Gaussian analysis likelihood, since it overlaps with the forecasted individual error (black circles). These forecasts gives good agreement with the calculated ensemble errors, therefore we can show that the underestimate of the Poisson errors does indeed appear to be due to lacking the sample variance contribution to the likelihood. 

\begin{figure*}
    \centering
    \includegraphics[width=.9\textwidth]{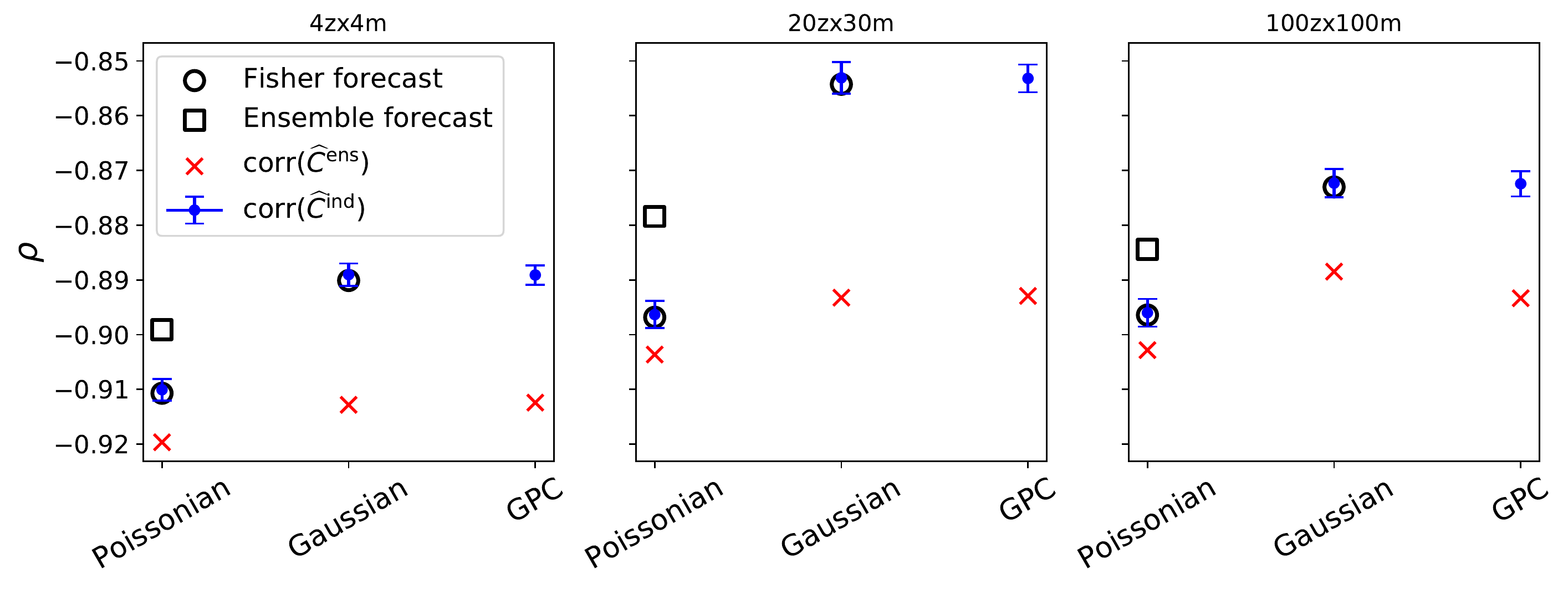}
    \caption{Correlation coefficients between $\Omega_m$ and $\sigma_8$ for the different binning setups in \cref{tab:binnings} (left to right). For each plot, the chosen likelihood is indicated on the x-axis while correlation is on the y-axis. Blue dots correspond to the average of individual correlations, and the error-bars represent to the standard deviation of these individual correlations}. Red crosses correspond to the correlation of the 1000 two-dimensional best fits. Black circles correspond to Fisher forecasts in \cref{eq:Fisher_formula_gaussian} (for the Poissonian and Gaussian likelihoods) and black boxes to the ensemble forecast in \cref{eq:cov_real_formula} (only for the Poisson likelihood, see text for details).
    \label{fig:rho_full}
    \centering
\end{figure*}

In addition to the errors on $\Omega_m$ and $\sigma_8$, we also compare the correlation coefficients using the three likelihoods and three binnings and present the results in \cref{fig:rho_full}. For each binning setup we find a negative correlation coefficient between the two cosmological parameters, which is the expected degeneracy between $\Omega_m$ and $\sigma_8$ with cluster abundance cosmology. Additionally we compute the correlation from the Fisher forecast and the ensemble forecasts. We find that the correlation is accurate for the Poisson likelihood, the blue dots are close to the red crosses. The correlation is slightly under-estimated for the GPC and Gaussian likelihoods (at most on the order of 5\% in the case of {\tt medium} binning). This is likely because only the block-diagonal terms are used in our covariance matrices for these likelihoods. This correlation would be important when calculating $S_8= \sigma_8 \sqrt{\Omega_m/0.3}$, which is often used to quantify the tension between CMB and large-scale structure cosmological probes (see \cite{2021Valentino} for a recent review). A 5\% bias on the correlation coefficient, like we see for the GPC and Gaussian cases, would lead to a bias of 1\% on the error of $S_8$ and is therefore unimportant.

Finally, in the three binning regimes where the tests above were undertaken, the Gaussian and GPC likelihood always obtain the same parameter constraints. The GPC provides, in principle, a more accurate description of the statistical behaviour of the data, but the Gaussian approximation is clearly sufficient in the three use cases. 

\begin{figure*}
    \centering
    \includegraphics[width=.7\textwidth]{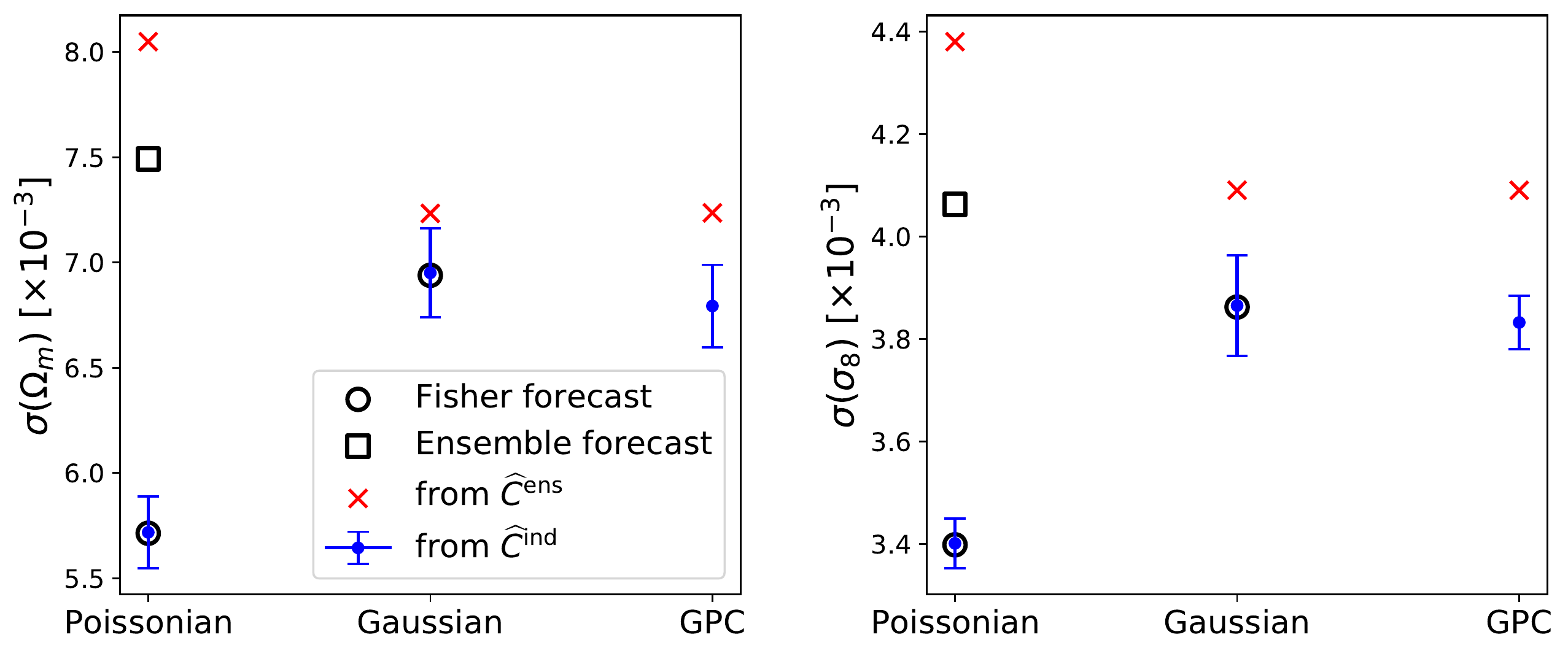}
    \includegraphics[width=.7\textwidth]{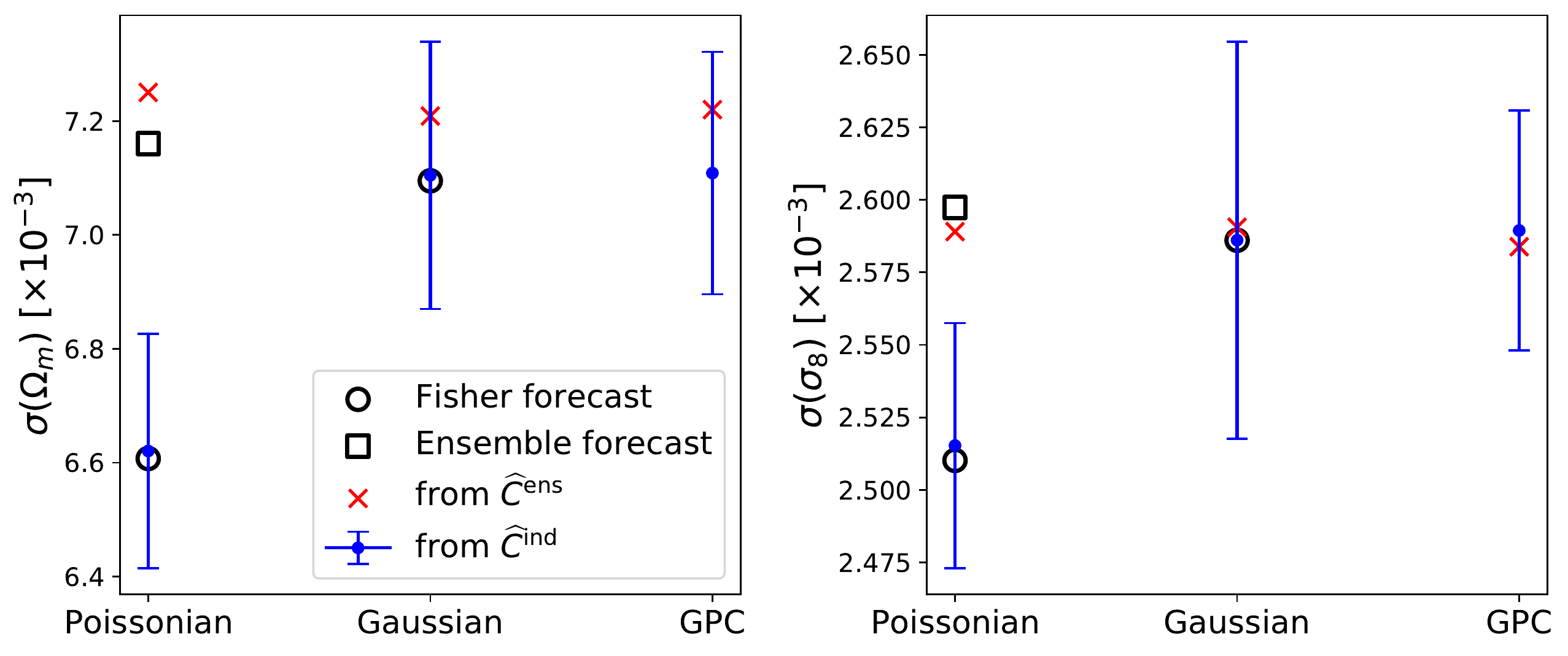}
    \caption{Same as \cref{fig:std_full} but for the reduced volume sample (top,  as described in \cref{subsec:partial}) and the high mass only cluster sample (bottom, as described in \cref{subsec:highmass}).}
    \label{fig:100zx100x_discussion}

\end{figure*}

The Gaussian likelihood remains an accurate description of the data, up to a very large number of bins, when the cluster analysis is performed with clusters above $10^{14}$~M$_\odot$ in a Rubin or {\it Euclid}-like survey. This is despite the bins being poorly populated. We would have expected that at a sufficiently fine binning that the Gaussian likelihood would no longer be valid, which we do not see in our results. But even with the \texttt{fine} binning, there are still a significant number of bins containing $\sim 100$ halos (see right panel in Fig.~\ref{fig:binning_scheme}). The answer then is that if we chosen an even finer binning that we would reach a regime in which the Gaussian approximation was no longer valid. However as we see little improvement between the cosmological constraints from 600 to 10,000 bins an even finer binning than that used appears to be of little cosmological interest.

\subsection{Reduced volume sample}
\label{subsec:partial}

As shown in \cref{section:cluster_distributions}, the Gaussian and Poisson likelihoods are two limit cases of the GPC description. In the previous section, where the number of clusters per bin is large, the GPC and Gaussian descriptions behave similarly and give robust constraints, conversely to the Poisson case. In this section and the next, we perform the following exercise: we reduce the number of clusters per bin to see if we can reach a regime where GPC and Poisson would give similar results and if those results could be more robust than the Gaussian case. To do so, we perform here another analysis in which we reduce the sky fraction by a factor 10, therefore reducing the total number of halo to $\sim 10^4$ per simulation. 

We show the results of the three likelihoods to this reduced volume sample using the {\tt fine} binning in \cref{fig:100zx100x_discussion}. By reducing the sky fraction, errors on $\Omega_m$ and $\sigma_8$ increased by a factor of $\sim 1/f_{\rm sky}$ compared to the constraints obtained using the full PINOCCHIO sky area in \cref{subsec:cov_robustness_measure}. The Poisson likelihood still underestimates errors by $\sim 25-30 \%$, and the Gaussian and GPC individual errors are closer to the ensemble errors, but negatively biased by a factor $\sim 5\%$.

\subsection{Considering high mass clusters only}
\label{subsec:highmass}

In the same spirit of \cref{subsec:partial}, we now reduce the number of objects by selecting high mass clusters only by selecting $M>5.10^{14} M_\odot$, but keeping the full PINOCCHIO sky area. We repeat the analysis with 100 mass bins this from $5.10^{14} M_\odot$ to $10^{15.6}\ M_\odot$ (we keep the upper limit is the same as defined in \cref{section:results}), and 100 redshift bins defined in \cref{tab:binnings}. The total number of halos used in the analysis now drops to $\sim 5\times 10^3$. We show in \cref{fig:100zx100x_discussion} (second line) the results of this work. We see that the Poisson likelihood still underestimates error on cosmological parameters from $5 - 20 \%$ (depending on the cosmological parameter). With this result, we see that off-diagonal terms of the sample covariance and the sample variance are still important. The Gaussian and GPC individual errors are fully consistent with the ensemble errors. In this regime and under our model assumptions, Gaussian and GPC describe correctly the latent likelihood.

\section{Conclusions}
\label{sec:conclusions}

Cluster abundance cosmology relies on a likelihood to extract cosmological information from the data that has been assumed in the past to follow a Poisson, Gaussian or the Gauss-Poisson Compound distribution. In the context of the forthcoming cluster surveys, such as the Rubin Legacy Survey of Space and Time or the {\it Euclid} survey, it is important to consider if these likelihoods are sufficiently accurate and robust.

To do so, we make use of the PINOCCHIO simulation set, that provides us with 1000 realisations of 10,313~$\deg^{2}$ (i.e. 1/4th off the full sky), for a given cosmology. For each likelihood and for each simulation, we perform a cosmological analysis using importance sampling to recover posteriors for $\sigma_8$ and $\Omega_m$. The comparison of the average and covariance of the 1000 posterior means to that of the individual posteriors, provides us with the metric to assess the reliability of the likelihood under scrutiny. This is repeated for three binning schemes in the mass-redshift plane, ranging from 16 to $10^4$ bins. Therefore in this work we have sampled a total of 9000 posteriors.

For each binning and each likelihood, we find a consistent bias in the recovered parameters despite this idealized set up. This bias is small with respect to the individual posterior variance and will therefore be negligible for future cosmological analyses. 

For the three binning schemes we have explored, over the full survey area, the Poisson likelihood always underestimates the errors on the parameters by $\sim 30\%$; this is unsurprising as the sample variance cannot be accounted for in that description. Reducing the sky fraction to $\sim 1030 ~\deg^{2}$ or limiting the analysis to the most massive clusters does not change that conclusion despite yielding bins populated by less than five clusters. The Gaussian and Gauss-Poisson Compound likelihoods perform similarly in all cases, underestimating the errors on the recovered parameters by at most 5\%.

Increasing the number of bins from the {\tt coarse} (16 bins) to {\tt fine} ($10^4$ bins) cases only yields a marginal $\sim 10\%$ improvement on $\sigma_8$ and $\Omega_m$ error bars. The advantage of a finer binning may therefore disappear in a real analysis when using a Gaussian likelihood \citep{Penna_Lima_2014,artis_2021,sartoris}. This may become clearer in the future as we are also performing a study about the impact of binned versus unbinned cluster count likelihoods on cosmological constraints (complementary to this work), focusing on the calibration of mass-proxy relations. Unbinned cluster count likelihoods and finer binnings may also be useful when studying the correlations between different observable properties of galaxy clusters; such as the cluster richness, weak lensing mass, X-ray mass and SZ masses (for example see \citealp{rozo2009,mahdavi2013,umetsu2020,murray2022}).

This analysis is also the opportunity to revisit the question of forecasts. We compute the Fisher forecasts for the Poisson and Gaussian likelihoods, finding them (as expected) in agreement with the individual posterior covariances. We also develop a forecast approach that, conversely to the Fisher forecast, allows us to account for a latent likelihood that is not necessarily the same as the {\it analysis} likelihood. In our work this allowed us to show that the difference between the Poisson individual posterior errors and the ensemble errors can be explained by the non-inclusion of the sample covariance in the Poisson likelihood. 

In more general terms, this new type of forecast, which in this work we call the ensemble forecast, would allow us to assess the reduction of the constraining power of an approximate likelihood in the case where the latent likelihood is known. This would be useful where the latent likelihood is computationally expensive to sample, for example, with many off-diagonal terms in the covariance matrix.

To conclude, from this work and for the general purpose of cluster abundance cosmology with future large scale surveys, the Gaussian likelihood remains a sufficient description of the data and we find no clear benefit to make the analysis more complex and slower using the GPC description or a large number of bins. 

\section*{Acknowledgements}
The authors thank the anonymous reviewer for their insightful comments and suggestions. We thank Pierluigi Monaco and Alex Saro for providing the PINOCCHIO catalogs, discussions of the work and providing useful comments on the manuscript. We thank Fabien Lacasa for his help in using the \texttt{PySSC} code. CM acknowledges funding from the French "Programme d’investissements d’avenir" through the Enigmass Labex. We gratefully acknowledge support from the CNRS/IN2P3 Computing Center (Lyon - France) for providing computing and data-processing resources needed for this work. 

\section*{Data availability}
The PINOCCHIO simulations used in this article cannot be shared publicly. The data will be shared on reasonable request to Alessandra Fumagalli (alessandra.fumagalli@inaf.it).

\bibliographystyle{mnras}
\bibliography{main}

\begin{thebibliography}{}
\makeatletter
\relax
\def\mn@urlcharsother{\let\do\@makeother \do\$\do\&\do\#\do\^\do\_\do\%\do\~}
\def\mn@doi{\begingroup\mn@urlcharsother \@ifnextchar [ {\mn@doi@}
  {\mn@doi@[]}}
\def\mn@doi@[#1]#2{\def\@tempa{#1}\ifx\@tempa\@empty \href
  {http://dx.doi.org/#2} {doi:#2}\else \href {http://dx.doi.org/#2} {#1}\fi
  \endgroup}
\def\mn@eprint#1#2{\mn@eprint@#1:#2::\@nil}
\def\mn@eprint@arXiv#1{\href {http://arxiv.org/abs/#1} {{\tt arXiv:#1}}}
\def\mn@eprint@dblp#1{\href {http://dblp.uni-trier.de/rec/bibtex/#1.xml}
  {dblp:#1}}
\def\mn@eprint@#1:#2:#3:#4\@nil{\def\@tempa {#1}\def\@tempb {#2}\def\@tempc
  {#3}\ifx \@tempc \@empty \let \@tempc \@tempb \let \@tempb \@tempa \fi \ifx
  \@tempb \@empty \def\@tempb {arXiv}\fi \@ifundefined
  {mn@eprint@\@tempb}{\@tempb:\@tempc}{\expandafter \expandafter \csname
  mn@eprint@\@tempb\endcsname \expandafter{\@tempc}}}

\bibitem[\protect\citeauthoryear{{Abbott} et~al.,}{{Abbott}
  et~al.}{2020}]{abbott2020dark}
{Abbott} T.~M.~C.,  et~al., 2020, \mn@doi [\prd] {10.1103/PhysRevD.102.023509},
  \href {https://ui.adsabs.harvard.edu/abs/2020PhRvD.102b3509A} {102, 023509}

\bibitem[\protect\citeauthoryear{{Artis}, {Melin}, {Bartlett}  \&
  {Murray}}{{Artis} et~al.}{2021}]{artis_2021}
{Artis} E.,  {Melin} J.-B.,  {Bartlett} J.~G.,   {Murray} C.,  2021, \mn@doi
  [\aap] {10.1051/0004-6361/202140293}, \href
  {https://ui.adsabs.harvard.edu/abs/2021A&A...649A..47A} {649, A47}

\bibitem[\protect\citeauthoryear{{Bocquet} et~al.,}{{Bocquet}
  et~al.}{2019}]{bocquet2019cluster}
{Bocquet} S.,  et~al., 2019, \mn@doi [\apj] {10.3847/1538-4357/ab1f10}, \href
  {https://ui.adsabs.harvard.edu/abs/2019ApJ...878...55B} {878, 55}

\bibitem[\protect\citeauthoryear{{Bond} \& {Myers}}{{Bond} \&
  {Myers}}{1996}]{bond1996peak}
{Bond} J.~R.,  {Myers} S.~T.,  1996, \mn@doi [\apjs] {10.1086/192267}, \href
  {https://ui.adsabs.harvard.edu/abs/1996ApJS..103....1B} {103, 1}

\bibitem[\protect\citeauthoryear{{Bond}, {Cole}, {Efstathiou}  \&
  {Kaiser}}{{Bond} et~al.}{1991}]{bond1991excursion}
{Bond} J.~R.,  {Cole} S.,  {Efstathiou} G.,   {Kaiser} N.,  1991, \mn@doi
  [\apj] {10.1086/170520}, \href
  {https://ui.adsabs.harvard.edu/abs/1991ApJ...379..440B} {379, 440}

\bibitem[\protect\citeauthoryear{{Bouchet}, {Colombi}, {Hivon}  \&
  {Juszkiewicz}}{{Bouchet} et~al.}{1995}]{bouchet1994perturbative}
{Bouchet} F.~R.,  {Colombi} S.,  {Hivon} E.,   {Juszkiewicz} R.,  1995, \aap,
  \href {https://ui.adsabs.harvard.edu/abs/1995A&A...296..575B} {296, 575}

\bibitem[\protect\citeauthoryear{{Buchert}}{{Buchert}}{1992}]{buchert1992lagrangian}
{Buchert} T.,  1992, \mn@doi [\mnras] {10.1093/mnras/254.4.729}, \href
  {https://ui.adsabs.harvard.edu/abs/1992MNRAS.254..729B} {254, 729}

\bibitem[\protect\citeauthoryear{Chisari et~al.,}{Chisari
  et~al.}{2019}]{ChisariCCL2019}
Chisari N.~E.,  et~al., 2019, \mn@doi [The Astrophysical Journal Supplement
  Series] {10.3847/1538-4365/ab1658}, 242, 2

\bibitem[\protect\citeauthoryear{{Cooray} \& {Sheth}}{{Cooray} \&
  {Sheth}}{2002}]{cooray2002halo}
{Cooray} A.,  {Sheth} R.,  2002, \mn@doi [\physrep]
  {10.1016/S0370-1573(02)00276-4}, \href
  {https://ui.adsabs.harvard.edu/abs/2002PhR...372....1C} {372, 1}

\bibitem[\protect\citeauthoryear{{Cramer}}{{Cramer}}{1952}]{CRRau}
{Cramer} R.,  1952, {Advanced statistical methods in biometric research}.
John Wiley $\&$ Sons, Inc.

\bibitem[\protect\citeauthoryear{{Despali}, {Giocoli}, {Angulo}, {Tormen},
  {Sheth}, {Baso}  \& {Moscardini}}{{Despali}
  et~al.}{2016}]{despali2016universality}
{Despali} G.,  {Giocoli} C.,  {Angulo} R.~E.,  {Tormen} G.,  {Sheth} R.~K.,
  {Baso} G.,   {Moscardini} L.,  2016, \mn@doi [\mnras]
  {10.1093/mnras/stv2842}, \href
  {https://ui.adsabs.harvard.edu/abs/2016MNRAS.456.2486D} {456, 2486}

\bibitem[\protect\citeauthoryear{{Di Valentino} et~al.,}{{Di Valentino}
  et~al.}{2021}]{2021Valentino}
{Di Valentino} E.,  et~al., 2021, \mn@doi [Astroparticle Physics]
  {10.1016/j.astropartphys.2021.102604}, \href
  {https://ui.adsabs.harvard.edu/abs/2021APh...13102604D} {131, 102604}

\bibitem[\protect\citeauthoryear{{Dodelson} \& {Schneider}}{{Dodelson} \&
  {Schneider}}{2013}]{dodelson2013effect}
{Dodelson} S.,  {Schneider} M.~D.,  2013, \mn@doi [\prd]
  {10.1103/PhysRevD.88.063537}, \href
  {https://ui.adsabs.harvard.edu/abs/2013PhRvD..88f3537D} {88, 063537}

\bibitem[\protect\citeauthoryear{Efron}{Efron}{2015}]{Efron}
Efron B.,  2015, \mn@doi [Journal of the Royal Statistical Society]
  {https://doi.org/10.1111/rssb.12080}, 77, 617

\bibitem[\protect\citeauthoryear{{Eisenstein} \& {Loeb}}{{Eisenstein} \&
  {Loeb}}{1995}]{eisenstein1994analytical}
{Eisenstein} D.~J.,  {Loeb} A.,  1995, \mn@doi [\apj] {10.1086/175193}, \href
  {https://ui.adsabs.harvard.edu/abs/1995ApJ...439..520E} {439, 520}

\bibitem[\protect\citeauthoryear{{Elvira} \& {Martino}}{{Elvira} \&
  {Martino}}{2021}]{2021arXiv210205407E}
{Elvira} V.,  {Martino} L.,  2021, arXiv e-prints, \href
  {https://ui.adsabs.harvard.edu/abs/2021arXiv210205407E} {p. arXiv:2102.05407}

\bibitem[\protect\citeauthoryear{{Evrard} et~al.,}{{Evrard}
  et~al.}{2002}]{evrard2002}
{Evrard} A.~E.,  et~al., 2002, \mn@doi [\apj] {10.1086/340551}, \href
  {https://ui.adsabs.harvard.edu/abs/2002ApJ...573....7E} {573, 7}

\bibitem[\protect\citeauthoryear{Fisher}{Fisher}{1935}]{FISHER}
Fisher R.~A.,  1935, Journal of the royal statistical society, 98, 39

\bibitem[\protect\citeauthoryear{{Fumagalli} et~al.,}{{Fumagalli}
  et~al.}{2021}]{Fumagalli_2021}
{Fumagalli} A.,  et~al., 2021, \mn@doi [\aap] {10.1051/0004-6361/202140592},
  \href {https://ui.adsabs.harvard.edu/abs/2021A&A...652A..21F} {652, A21}

\bibitem[\protect\citeauthoryear{{Gouyou Beauchamps}, {Lacasa}, {Tutusaus},
  {Aubert}, {Baratta}, {Gorce}  \& {Sakr}}{{Gouyou Beauchamps}
  et~al.}{2022}]{2022A&A...659A.128G}
{Gouyou Beauchamps} S.,  {Lacasa} F.,  {Tutusaus} I.,  {Aubert} M.,  {Baratta}
  P.,  {Gorce} A.,   {Sakr} Z.,  2022, \mn@doi [\aap]
  {10.1051/0004-6361/202142052}, \href
  {https://ui.adsabs.harvard.edu/abs/2022A&A...659A.128G} {659, A128}

\bibitem[\protect\citeauthoryear{{Heavens}, {Seikel}, {Nord}, {Aich},
  {Bouffanais}, {Bassett}  \& {Hobson}}{{Heavens} et~al.}{2014}]{heavens_2014}
{Heavens} A.~F.,  {Seikel} M.,  {Nord} B.~D.,  {Aich} M.,  {Bouffanais} Y.,
  {Bassett} B.~A.,   {Hobson} M.~P.,  2014, \mn@doi [\mnras]
  {10.1093/mnras/stu1866}, \href
  {https://ui.adsabs.harvard.edu/abs/2014MNRAS.445.1687H} {445, 1687}

\bibitem[\protect\citeauthoryear{{Henson}, {Barnes}, {Kay}, {McCarthy}  \&
  {Schaye}}{{Henson} et~al.}{2017}]{henson2016impact}
{Henson} M.~A.,  {Barnes} D.~J.,  {Kay} S.~T.,  {McCarthy} I.~G.,   {Schaye}
  J.,  2017, \mn@doi [\mnras] {10.1093/mnras/stw2899}, \href
  {https://ui.adsabs.harvard.edu/abs/2017MNRAS.465.3361H} {465, 3361}

\bibitem[\protect\citeauthoryear{Hu \& Kravtsov}{Hu \&
  Kravtsov}{2003}]{2003_SV_HU}
Hu W.,  Kravtsov A.~V.,  2003, \mn@doi [The Astrophysical Journal]
  {10.1086/345846}, 584, 702–715

\bibitem[\protect\citeauthoryear{{Jenkins}, {Frenk}, {White}, {Colberg},
  {Cole}, {Evrard}, {Couchman}  \& {Yoshida}}{{Jenkins}
  et~al.}{2001}]{jenkins2001}
{Jenkins} A.,  {Frenk} C.~S.,  {White} S.~D.~M.,  {Colberg} J.~M.,  {Cole} S.,
  {Evrard} A.~E.,  {Couchman} H.~M.~P.,   {Yoshida} N.,  2001, \mn@doi [\mnras]
  {10.1046/j.1365-8711.2001.04029.x}, \href
  {https://ui.adsabs.harvard.edu/abs/2001MNRAS.321..372J} {321, 372}

\bibitem[\protect\citeauthoryear{{Kaiser}}{{Kaiser}}{1984}]{kaiser1984spatial}
{Kaiser} N.,  1984, \mn@doi [\apjl] {10.1086/184341}, \href
  {https://ui.adsabs.harvard.edu/abs/1984ApJ...284L...9K} {284, L9}

\bibitem[\protect\citeauthoryear{{Ku}}{{Ku}}{1996}]{errorprop}
{Ku} H.~H.,  1996, NIST, 70, 263

\bibitem[\protect\citeauthoryear{{LSST Science Collaboration} et~al.,}{{LSST
  Science Collaboration} et~al.}{2009}]{LSST}
{LSST Science Collaboration} et~al., 2009, arXiv e-prints, \href
  {https://ui.adsabs.harvard.edu/abs/2009arXiv0912.0201L} {p. arXiv:0912.0201}

\bibitem[\protect\citeauthoryear{Lacasa, Lima  \& Aguena}{Lacasa
  et~al.}{2018}]{2018_Lacasa_SSC_partial_sky}
Lacasa F.,  Lima M.,   Aguena M.,  2018, \mn@doi [Astronomy & Astrophysics]
  {10.1051/0004-6361/201630281}, 611, A83

\bibitem[\protect\citeauthoryear{{Laureijs} et~al.,}{{Laureijs}
  et~al.}{2011}]{laureijs2011euclid}
{Laureijs} R.,  et~al., 2011, arXiv e-prints, \href
  {https://ui.adsabs.harvard.edu/abs/2011arXiv1110.3193L} {p. arXiv:1110.3193}

\bibitem[\protect\citeauthoryear{{Lesci} et~al.,}{{Lesci}
  et~al.}{2022a}]{2022arXiv220307398L}
{Lesci} G.~F.,  et~al., 2022a, arXiv e-prints, \href
  {https://ui.adsabs.harvard.edu/abs/2022arXiv220307398L} {p. arXiv:2203.07398}

\bibitem[\protect\citeauthoryear{{Lesci} et~al.,}{{Lesci}
  et~al.}{2022b}]{lesci2022amico}
{Lesci} G.~F.,  et~al., 2022b, \mn@doi [\aap] {10.1051/0004-6361/202040194},
  \href {https://ui.adsabs.harvard.edu/abs/2022A&A...659A..88L} {659, A88}

\bibitem[\protect\citeauthoryear{Lima \& Hu}{Lima \& Hu}{2004}]{Lima_Hu}
Lima M.,  Hu W.,  2004, \mn@doi [Phys. Rev. D] {10.1103/PhysRevD.70.043504},
  70, 043504

\bibitem[\protect\citeauthoryear{{Mahdavi}, {Hoekstra}, {Babul}, {Bildfell},
  {Jeltema}  \& {Henry}}{{Mahdavi} et~al.}{2013}]{mahdavi2013}
{Mahdavi} A.,  {Hoekstra} H.,  {Babul} A.,  {Bildfell} C.,  {Jeltema} T.,
  {Henry} J.~P.,  2013, \mn@doi [\apj] {10.1088/0004-637X/767/2/116}, \href
  {https://ui.adsabs.harvard.edu/abs/2013ApJ...767..116M} {767, 116}

\bibitem[\protect\citeauthoryear{{Mantz} et~al.,}{{Mantz}
  et~al.}{2015}]{mantz2015weighing}
{Mantz} A.~B.,  et~al., 2015, \mn@doi [\mnras] {10.1093/mnras/stu2096}, \href
  {https://ui.adsabs.harvard.edu/abs/2015MNRAS.446.2205M} {446, 2205}

\bibitem[\protect\citeauthoryear{{Monaco}}{{Monaco}}{1997}]{Monaco1997}
{Monaco} P.,  1997, \mn@doi [\mnras] {10.1093/mnras/287.4.753}, \href
  {https://ui.adsabs.harvard.edu/abs/1997MNRAS.287..753M} {287, 753}

\bibitem[\protect\citeauthoryear{{Monaco}, {Theuns}  \& {Taffoni}}{{Monaco}
  et~al.}{2002}]{monaco2002pinocchio}
{Monaco} P.,  {Theuns} T.,   {Taffoni} G.,  2002, \mn@doi [\mnras]
  {10.1046/j.1365-8711.2002.05162.x}, \href
  {https://ui.adsabs.harvard.edu/abs/2002MNRAS.331..587M} {331, 587}

\bibitem[\protect\citeauthoryear{{Moutarde}, {Alimi}, {Bouchet}, {Pellat}  \&
  {Ramani}}{{Moutarde} et~al.}{1991}]{moutarde1991precollapse}
{Moutarde} F.,  {Alimi} J.~M.,  {Bouchet} F.~R.,  {Pellat} R.,   {Ramani} A.,
  1991, \mn@doi [\apj] {10.1086/170728}, \href
  {https://ui.adsabs.harvard.edu/abs/1991ApJ...382..377M} {382, 377}

\bibitem[\protect\citeauthoryear{{Munari}, {Monaco}, {Sefusatti}, {Castorina},
  {Mohammad}, {Anselmi}  \& {Borgani}}{{Munari}
  et~al.}{2017}]{munari2017improving}
{Munari} E.,  {Monaco} P.,  {Sefusatti} E.,  {Castorina} E.,  {Mohammad} F.~G.,
   {Anselmi} S.,   {Borgani} S.,  2017, \mn@doi [\mnras]
  {10.1093/mnras/stw3085}, \href
  {https://ui.adsabs.harvard.edu/abs/2017MNRAS.465.4658M} {465, 4658}

\bibitem[\protect\citeauthoryear{{Murray}, {Bartlett}, {Artis}  \&
  {Melin}}{{Murray} et~al.}{2022}]{murray2022}
{Murray} C.,  {Bartlett} J.~G.,  {Artis} E.,   {Melin} J.-B.,  2022, \mn@doi
  [\mnras] {10.1093/mnras/stac689}, \href
  {https://ui.adsabs.harvard.edu/abs/2022MNRAS.512.4785M} {512, 4785}

\bibitem[\protect\citeauthoryear{{Penna-Lima}, {Makler}  \&
  {Wuensche}}{{Penna-Lima} et~al.}{2014}]{Penna_Lima_2014}
{Penna-Lima} M.,  {Makler} M.,   {Wuensche} C.~A.,  2014, \mn@doi [\jcap]
  {10.1088/1475-7516/2014/05/039}, \href
  {https://ui.adsabs.harvard.edu/abs/2014JCAP...05..039P} {2014, 039}

\bibitem[\protect\citeauthoryear{{Percival}, {Friedrich}, {Sellentin}  \&
  {Heavens}}{{Percival} et~al.}{2022}]{percival2022matching}
{Percival} W.~J.,  {Friedrich} O.,  {Sellentin} E.,   {Heavens} A.,  2022,
  \mn@doi [\mnras] {10.1093/mnras/stab3540}, \href
  {https://ui.adsabs.harvard.edu/abs/2022MNRAS.510.3207P} {510, 3207}

\bibitem[\protect\citeauthoryear{{Planck Collaboration} et~al.,}{{Planck
  Collaboration} et~al.}{2014}]{Planck_14}
{Planck Collaboration} et~al., 2014, \mn@doi [\aap]
  {10.1051/0004-6361/201321591}, \href
  {https://ui.adsabs.harvard.edu/abs/2014A&A...571A..16P} {571, A16}

\bibitem[\protect\citeauthoryear{{Planck Collaboration} et~al.,}{{Planck
  Collaboration} et~al.}{2016}]{ade2016planck}
{Planck Collaboration} et~al., 2016, \mn@doi [\aap]
  {10.1051/0004-6361/201525833}, \href
  {https://ui.adsabs.harvard.edu/abs/2016A&A...594A..24P} {594, A24}

\bibitem[\protect\citeauthoryear{{Press} \& {Schechter}}{{Press} \&
  {Schechter}}{1974}]{press1974formation}
{Press} W.~H.,  {Schechter} P.,  1974, \mn@doi [\apj] {10.1086/152650}, \href
  {https://ui.adsabs.harvard.edu/abs/1974ApJ...187..425P} {187, 425}

\bibitem[\protect\citeauthoryear{Raveri \& Hu}{Raveri \&
  Hu}{2019}]{Raveri_2019}
Raveri M.,  Hu W.,  2019, \mn@doi [Physical Review D]
  {10.1103/physrevd.99.043506}, 99

\bibitem[\protect\citeauthoryear{{Rozo} et~al.,}{{Rozo}
  et~al.}{2009}]{rozo2009}
{Rozo} E.,  et~al., 2009, \mn@doi [\apj] {10.1088/0004-637X/699/1/768}, \href
  {https://ui.adsabs.harvard.edu/abs/2009ApJ...699..768R} {699, 768}

\bibitem[\protect\citeauthoryear{{Sartoris} et~al.,}{{Sartoris}
  et~al.}{2016}]{sartoris}
{Sartoris} B.,  et~al., 2016, \mn@doi [\mnras] {10.1093/mnras/stw630}, \href
  {https://ui.adsabs.harvard.edu/abs/2016MNRAS.459.1764S} {459, 1764}

\bibitem[\protect\citeauthoryear{{Sellentin} \& {Heavens}}{{Sellentin} \&
  {Heavens}}{2017}]{sellentin2016quantifying}
{Sellentin} E.,  {Heavens} A.~F.,  2017, \mn@doi [\mnras]
  {10.1093/mnras/stw2697}, \href
  {https://ui.adsabs.harvard.edu/abs/2017MNRAS.464.4658S} {464, 4658}

\bibitem[\protect\citeauthoryear{{Sellentin} \& {Heavens}}{{Sellentin} \&
  {Heavens}}{2018}]{sellentin2018insufficiency}
{Sellentin} E.,  {Heavens} A.~F.,  2018, \mn@doi [\mnras]
  {10.1093/mnras/stx2491}, \href
  {https://ui.adsabs.harvard.edu/abs/2018MNRAS.473.2355S} {473, 2355}

\bibitem[\protect\citeauthoryear{{Sellentin}, {Quartin}  \&
  {Amendola}}{{Sellentin} et~al.}{2014}]{Sellentinforecast14}
{Sellentin} E.,  {Quartin} M.,   {Amendola} L.,  2014, \mn@doi [\mnras]
  {10.1093/mnras/stu689}, \href
  {https://ui.adsabs.harvard.edu/abs/2014MNRAS.441.1831S} {441, 1831}

\bibitem[\protect\citeauthoryear{{Sheth} \& {Tormen}}{{Sheth} \&
  {Tormen}}{1999}]{sheth_tormen}
{Sheth} R.~K.,  {Tormen} G.,  1999, \mn@doi [\mnras]
  {10.1046/j.1365-8711.1999.02692.x}, \href
  {https://ui.adsabs.harvard.edu/abs/1999MNRAS.308..119S} {308, 119}

\bibitem[\protect\citeauthoryear{Takada \& Spergel}{Takada \&
  Spergel}{2014}]{2014_Takada}
Takada M.,  Spergel D.~N.,  2014, \mn@doi [Monthly Notices of the Royal
  Astronomical Society] {10.1093/mnras/stu759}, 441, 2456–2475

\bibitem[\protect\citeauthoryear{{Taylor} \& {Joachimi}}{{Taylor} \&
  {Joachimi}}{2014}]{taylor2014estimating}
{Taylor} A.,  {Joachimi} B.,  2014, \mn@doi [\mnras] {10.1093/mnras/stu996},
  \href {https://ui.adsabs.harvard.edu/abs/2014MNRAS.442.2728T} {442, 2728}

\bibitem[\protect\citeauthoryear{{Tegmark}, {Taylor}  \& {Heavens}}{{Tegmark}
  et~al.}{1997}]{tegmark}
{Tegmark} M.,  {Taylor} A.~N.,   {Heavens} A.~F.,  1997, \mn@doi [\apj]
  {10.1086/303939}, \href
  {https://ui.adsabs.harvard.edu/abs/1997ApJ...480...22T} {480, 22}

\bibitem[\protect\citeauthoryear{{Tinker}, {Kravtsov}, {Klypin}, {Abazajian},
  {Warren}, {Yepes}, {Gottl{\"o}ber}  \& {Holz}}{{Tinker}
  et~al.}{2008}]{tinker2008}
{Tinker} J.,  {Kravtsov} A.~V.,  {Klypin} A.,  {Abazajian} K.,  {Warren} M.,
  {Yepes} G.,  {Gottl{\"o}ber} S.,   {Holz} D.~E.,  2008, \mn@doi [\apj]
  {10.1086/591439}, \href
  {https://ui.adsabs.harvard.edu/abs/2008ApJ...688..709T} {688, 709}

\bibitem[\protect\citeauthoryear{Tinker, Robertson, Kravtsov, Klypin, Warren,
  Yepes  \& Gottlöber}{Tinker et~al.}{2010}]{2010_tinker}
Tinker J.~L.,  Robertson B.~E.,  Kravtsov A.~V.,  Klypin A.,  Warren M.~S.,
  Yepes G.,   Gottlöber S.,  2010, \mn@doi [The Astrophysical Journal]
  {10.1088/0004-637x/724/2/878}, 724, 878–886

\bibitem[\protect\citeauthoryear{{Umetsu} et~al.,}{{Umetsu}
  et~al.}{2020}]{umetsu2020}
{Umetsu} K.,  et~al., 2020, \mn@doi [\apj] {10.3847/1538-4357/ab6bca}, \href
  {https://ui.adsabs.harvard.edu/abs/2020ApJ...890..148U} {890, 148}

\bibitem[\protect\citeauthoryear{Wolz, Kilbinger, Weller  \& Giannantonio}{Wolz
  et~al.}{2012}]{Wolz_2012}
Wolz L.,  Kilbinger M.,  Weller J.,   Giannantonio T.,  2012, \mn@doi [Journal
  of Cosmology and Astroparticle Physics] {10.1088/1475-7516/2012/09/009},
  2012, 009

\bibitem[\protect\citeauthoryear{{Zwicky}}{{Zwicky}}{1933}]{zwicky}
{Zwicky} F.,  1933, Helvetica Physica Acta, \href
  {https://ui.adsabs.harvard.edu/abs/1933AcHPh...6..110Z} {6, 110}

\makeatother
\end{thebibliography}

\appendix

\section{Sample variance with PySSC}
\label{app:sample_variance_pyssc}

The sample covariance $\Sigma_{\rm SV}$ between two counts $N_{ij}$ and $N_{kl}$ is given by \citep{2003_SV_HU}
\begin{equation}
    \Sigma_{\rm SV}[ij][kl] =  \int_{z_j}^{z_{j+1}}dV_1\int_{z_l}^{z_{l+1}}dV_2\frac{\partial \mathcal{O}_{i}(z_1)}{\partial \delta}\frac{\partial \mathcal{O}_{k}(z_2)}{\partial \delta}\sigma^2(z_1, z_2)\;,
\end{equation}
where the quantity 
\begin{equation}
    \frac{\partial \mathcal{O}_{i}(z)}{\partial \delta} = \int_{m_i}^{m_{i+1}} dm b(m,z) \frac{dn(m,z)}{dm}
\end{equation}
is the \textit{response} of the probe $\mathcal{O}_{i}$ defined as
\begin{equation}
    \mathcal{O}_{i}(z) = \int_{m_i}^{m_{i+1}}dm\frac{dn(m,z)}{dm}( 1 + b(m,z)\delta)
\end{equation}
relative to the background density field. The quantity $\sigma^2(z_1, z_2)$ is the amplitude fluctuation of the matter power spectrum between two different redshifts and depends on the survey window function. Considering the full sky area, $\sigma^2(z_1, z_2)$ is defined as
\begin{equation}
    \sigma^2(z_1, z_2) = \frac{1}{2\pi^2}\int k^2 dk j_0(kw(z_1))j_0(kw(z_2))P_{m}(k|z_1, z_2),
\end{equation}
where $w$ is the line-of-sight comoving distance, and $j_0$ is the spherical Bessel function of order 0, $P_{m}(k|z_1, z_2)$ is the matter power spectrum at redshift $z_1$ and $z_2$.
Assuming the $\frac{\partial \mathcal{O}_{\alpha}(z)}{\partial \delta}$ varies slowly with redshift compared to the matter fluctuation amplitude, we get the approximation \citep{2018_Lacasa_SSC_partial_sky}
\begin{equation}
   \Sigma_{\rm SV}^{\rm th}[ij][kl] \approx \langle b\lambda\rangle_{ij} \langle b\lambda\rangle_{kl} S_{jl}\;,
\end{equation}
where $\langle b\lambda\rangle_{ij}$ is given in \cref{eq:Nbias}. The matrix $S_{jl}^{\rm fullsky}$ is the average amplitude fluctuation over the two separate volume and is defined by
\begin{equation}
    S_{jl}^{\rm fullsky} = \frac{1}{V_{j}V_{l}}\int_{z_j}^{z_{j+1}} dV_1\int_{z_l}^{z_{l+1}}dV_{2}\ \sigma^2(z_1, z_2).
    \label{eq:S_ij}
\end{equation}
In practice, we use the \texttt{PySSC} package to compute $S_{jl}^{\rm fullsky}$ matrix in \cref{eq:S_ij} \citep{2018_Lacasa_SSC_partial_sky}. We then rescale it by the sky fraction and use $ S_{jl} \approx S_{jl}^{\rm fullsky}/f_{\rm sky} = S_{jl}^{\rm fullsky}\times 4\pi/\Omega_S$ \citep{2018_Lacasa_SSC_partial_sky}. While this is only an approximation, \citet{2022A&A...659A.128G} shows that this performs well for sufficiently large sky areas, such as the one covered by the PINOCCHIO simulations. To check our use of \texttt{PySSC}, we compare the full theoretical prediction of the covariance, given by $\Sigma^{\rm th} = \Sigma_{\rm SV}^{\rm th} + \Sigma_{\rm SN}^{\rm th}$ ($\Sigma_{\rm SN}^{\rm th}$ is given in Eq. \eqref{eq:sn_theo}), to the covariance directly estimated from the data. The estimated full covariance $\widehat{\Sigma}^{\rm data}[ij][kl]$ between 2D mass-redshift bins $ij$ and $kl$ calculated from the 1000 simulations,

\begin{equation}
    \widehat{\Sigma}^{\rm data}[ij][kl] = \frac{1}{N_{\rm sim}-1}\sum_{n = 1}^{{N_{\rm sim}}}\left([\widehat{N}_{ij}]_n - \bar{N}_{ij}\right)\left([\widehat{N}_{kl}]_n - \bar{N}_{kl}\right),
    \label{eq:estimation_covariance}
\end{equation}

where $N_{\rm sim}$ is the number simulations, $[\widehat{N}_{ij}]_{n}$ is the measured number of clusters in bin $ij$ for the $n$-th PINOCCHIO simulation, and $\widehat{N}_{ij}$ is the cluster abundance measured in bin $ij$ average over the 1000 simulations.

\begin{figure*}
    \centering
    \includegraphics[width=.45\textwidth]{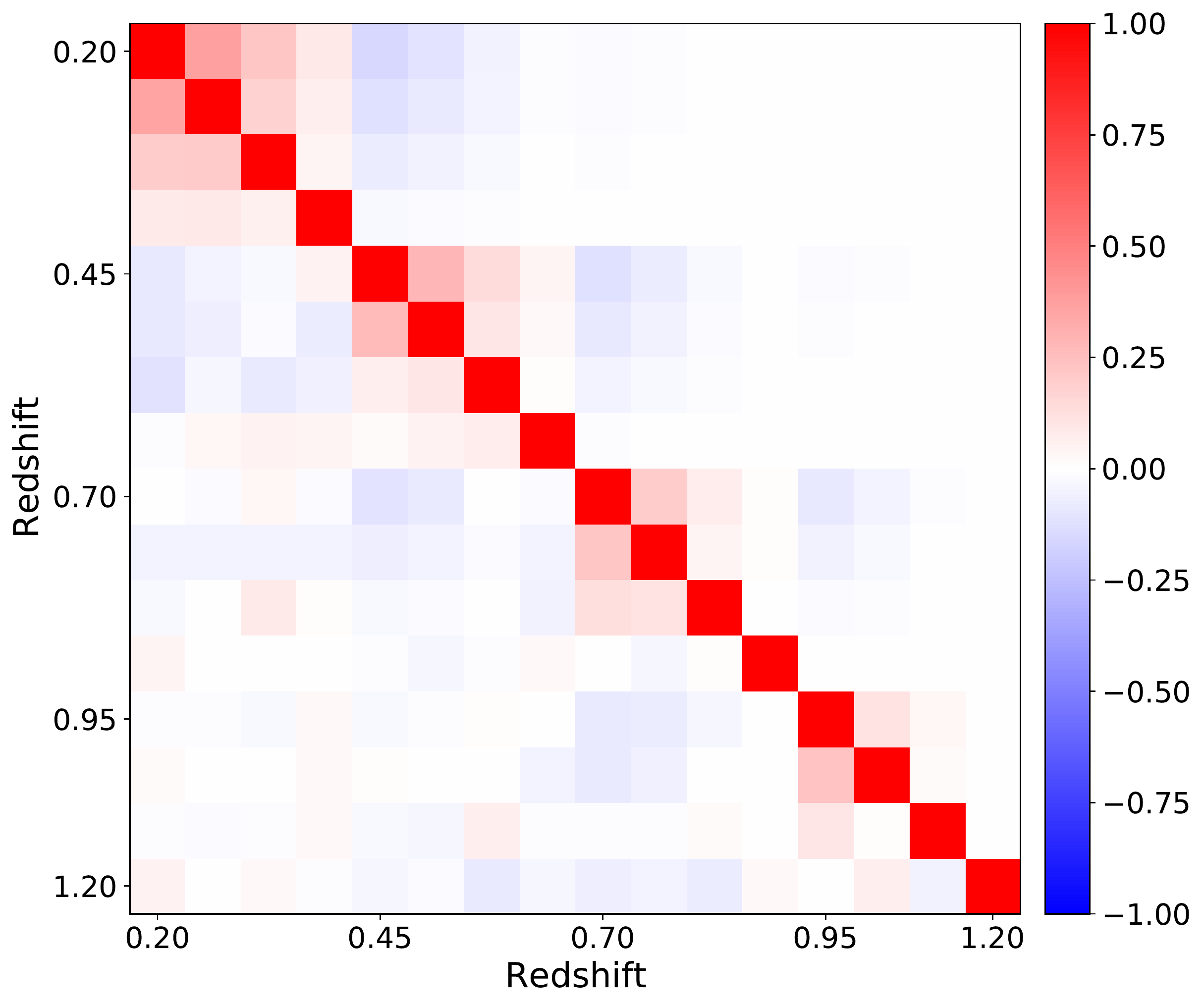}
    \hspace{.4cm}
    \includegraphics[width=.43\textwidth]{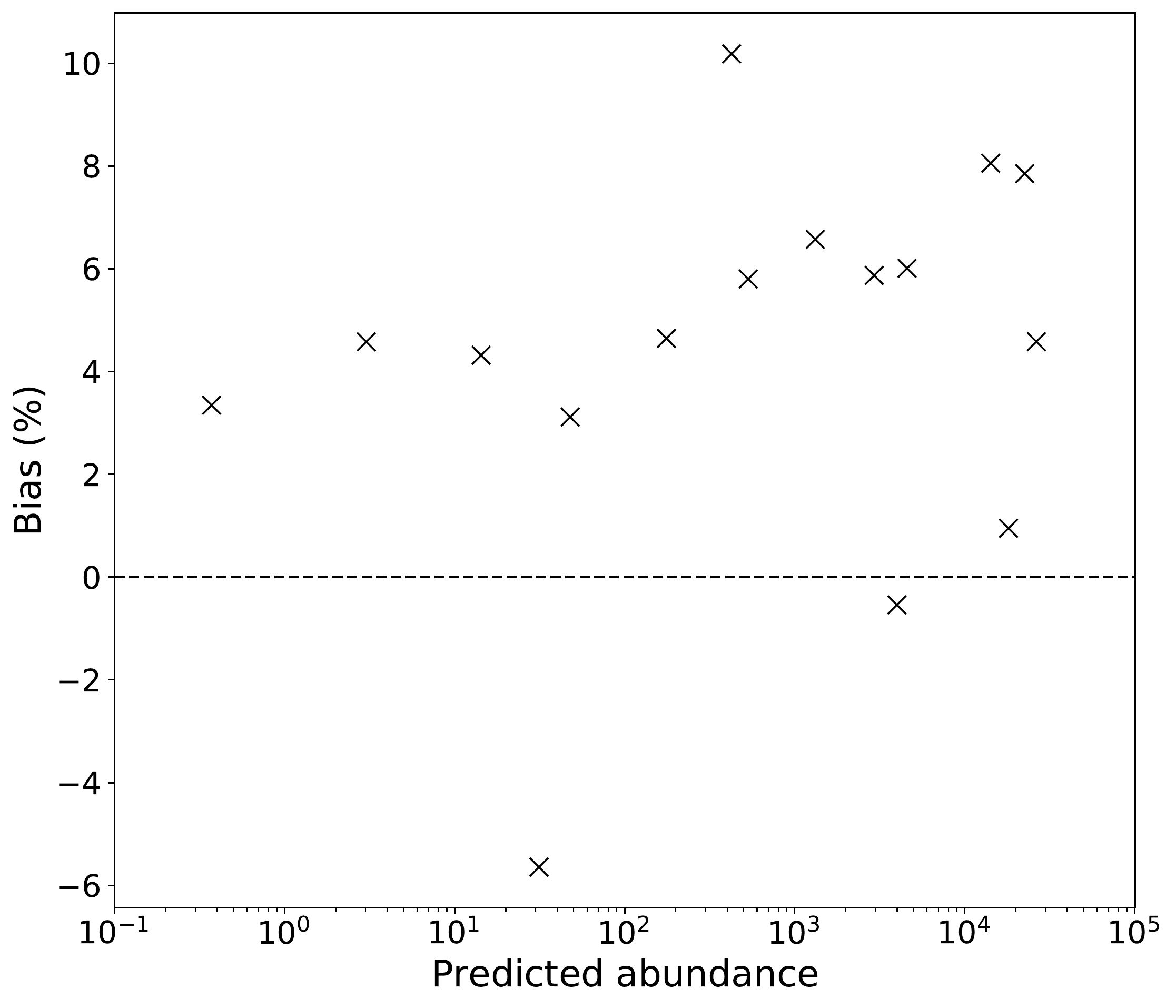}
    \caption{{\it Left}: Correlation matrix of cluster counts for the {\tt coarse} binning scheme. The upper triangle corresponds to the predicted correlation matrix from Eq. \eqref{eq:covariance} and the lower triangle to the estimated one from Eq. \eqref{eq:estimation_covariance}. {\it Right}: Relative bias of the predicted variance of cluster abundance relative to the estimated variance computed in \cref{eq:bias_variance_cluster_count} as a function of abundance $N$. \label{fig:bias_err}}
\end{figure*}

The left panel in \cref{fig:bias_err} shows the correlation matrix, defined as,
\begin{equation}
    R[ij][kl]=\frac{\Sigma[ij][kl]}{\sqrt{\Sigma[ij][ij]\Sigma[kl][kl]}}.
    \label{eq:correlation}
\end{equation}
for the data-derived covariance (lower triangle) and the theoretical covariance prediction (upper triangle). This is shown for the coarsest binning scheme (see \cref{subsec:binning} and \cref{fig:binning_scheme}) with the correlation matrix organised in four redshift bins, appearing as block diagonal, inside which the four mass bins are added.

We find the correlation between bins to be well-represented by the theoretical prediction for the block diagonal terms and the first off-diagonal terms. However, for more distant correlations, statistical noise dominates the data-derived covariance matrix (lower triangle) while the correlation drops to zero in the theoretical estimation (upper triangle). 

To further compare the two estimations, we compute the relative difference of the variance (diagonal elements) between the two approaches, defined as
\begin{equation}
    b_{ij} = \frac{\Sigma^{\rm data}[ij][ij]-\Sigma^{\rm th}[ij][ij]}{\Sigma^{\rm th}[ij][ij]}.
    \label{eq:bias_variance_cluster_count}
\end{equation}
The results are represented as a function of the mean abundance, for the 16 diagonal elements of the $4\times 4$ {\tt coarse} binning scheme, in the right panel in \cref{fig:bias_err}. We find a bias of about $10 \%$, which is expected according to the precision of PINOCCHIO two-point statistics.

This check validates our computation of the theoretical covariance, which we favour in our analysis as the data-derived covariance is affected by statistical noise due to the limited number of samples. 


\section{Variance of estimators}
\label{app:var_est}

In this appendix, we forecast the ensemble covariance matrix in Eq. \eqref{eq:cov_ensemble} in terms of both statistical properties of the data encoded in the input and latent likelihoods. We also show that the ensemble and individual covariance matrices are equivalent when the input and latent likelihoods are equivalent when we consider the simple case of Gaussian likelihoods.

From \citet{Efron}, we have that the gradient of the estimator $\nabla \widehat{\theta}_\alpha$ with respect to the data is given by
\begin{equation}
    [\nabla \widehat{\theta}_\alpha]_i = \mathrm{Cov}_{\mathcal{P}_Y}[\{\theta_\alpha, [\nabla \ln \mathcal{L}_Y(\bm{N}|\theta)]_i\}|\bm{N}].
    \label{eq:nabla_theta}
\end{equation}
From \cref{eq:nabla_theta}, we can derive the forecast for the ensemble covariance matrix, depending on the properties of the analysis likelihood. We first study the case of a Gaussian analysis likelihood, and the case of a Poisson likelihood.
\subsection{Gaussian analysis likelihood}
\label{app:var_est_1}
The gradient $\nabla \ln \mathcal{L}_Y$ can be expressed for the particular case of a Gaussian likelihood $\mathcal{L}_Y$ with an analysis covariance matrix $\Sigma_Y$, and we have that
\begin{equation}
    \nabla \ln \mathcal{L}_Y(\bm{N}|\theta)=\Sigma_Y^{-1}[\bm{\lambda}(\theta)-\bm{N}].
    \label{eq:nabla_lnL}
\end{equation}
Using Eq. \eqref{eq:nabla_lnL} in Eq. \eqref{eq:nabla_theta}, we get the general expression, which is valid for an \textit{input} Gaussian likelihood $\mathcal{L}_Y$ that
\begin{equation}
    [\nabla \widehat{\theta}_\alpha]_i = [\Sigma_Y^{-1}]_{ij}\mathrm{Cov}_{\mathcal{P}_Y}[\{\theta_\alpha, \lambda_j(\bm{\theta)}\}|\bm{N}].
    \label{eq:nabla_theta_2nd}
\end{equation}
Second, we consider the parameter estimator $\widehat{\theta}(\bm{N})$, at fixed observed cluster count $\bm{N}$. We now consider that the observed cluster count $\bm{N}$ corresponds to the \textit{latent} average cluster count $\lambda_0$ with small additive deviation, i.e. we can write $\bm{N} = \bm{\lambda}_0 + \bm{\delta N}$. In this formalism, we state that $\bm{\theta}_0$ are the \textit{latent} cosmological parameters giving $\bm{\lambda}_0 = \lambda(\bm{\theta}_0)$, and we state that $\bm{\lambda}_0 = \mathbb{E}_{\mathcal{L}_Y}(\bm{N}|\bm{\theta}_0) = \mathbb{E}_{\mathcal{L}_X}(\bm{N}|\bm{\theta}_0)$. Using the expression of the posterior in \cref{eq:Posterior}, of the estimator in \cref{eq:estimator_mean} and the posterior variance in \cref{eq:individual_covariance}, we consider the Taylor expansion of the estimator $\widehat{\theta}(\bm{\lambda}_0 + \bm{\delta N})$ considering the deviation $\bm{\delta N} = \bm{N} - \bm{\lambda}_0$, so we get for the first order that

\begin{equation}
    \widehat{\theta}_\alpha(\bm{N}) - \widehat{\theta}_\alpha(\lambda_0) \approx [\nabla \widehat{\theta}_\alpha]_i \cdot [\bm{N} - \bm{\lambda}_0]_i,
    \label{eq:e1}
\end{equation}

where in Eq. \eqref{eq:e1}, $[\nabla \widehat{\theta}_\alpha]_i$ corresponds to the gradient defined in Eq. \eqref{eq:nabla_theta} evaluated at $\bm{N} = \bm{\lambda}_0$. For \cref{eq:e1} to be valid, it is assumed that the estimator $\widehat{\theta}(\bm{N})$ is \textit{sufficiently} linear around $\bm{\lambda}_0$ with respect to the data, as it is considered for linear error propagation from data to estimated parameter \citep{errorprop}.  We now assume that the model is well approximated by a linear dependence on the parameter of interest, e.g. \citet{Raveri_2019}, and taking the first order in the Taylor expansion of $\lambda(\theta)$ at $\theta_0$, we get that

\begin{equation}
    [\nabla \widehat{\theta}_\alpha]_i = [\Sigma_Y^{-1}]_{ij}\mathcal{C}_{\alpha \gamma}^{\rm ind} (\bm{\lambda}_{,\gamma})_j =[\Sigma_Y^{-1}(\mathcal{C}_{\alpha \gamma}^{\rm ind} \bm{\lambda}_{,\gamma})]_i,
    \label{eq:nabla_theta_2}
\end{equation}

where $\mathcal{C}_{\alpha \gamma}^{\rm ind} = \mathrm{Cov}_{\mathcal{P}_Y}[\{\theta_\alpha, \theta_\beta\}|\bm{\lambda}_0]$ is the posterior covariance matrix as defined in \cref{eq:individual_covariance}, and $\bm{\lambda}_{,\gamma}$ is the derivative of the model prediction evaluated at $\theta = \theta_0$. Note that we subsequently conserved the first order moments of 2 Taylor expansions respectively in the data and parameter space, that need to be valid to follow the proof. We will consider these approximations correct for the following steps.
Re-writing the expression of the ensemble covariance in \cref{eq:cov_ensemble}, we find that

\begin{align}
\mathcal{C}_{\alpha\beta}^{\mathrm{ens}}&=\mathrm{Cov}_{\mathcal{L}_X} [\{\widehat{\theta}_\alpha(\bm{N}), \widehat{\theta}_\beta(\bm{N})\}|\bm{\theta}_0]\\
    &=(\mathcal{C}^{\mathrm{ind}} \bm{\lambda}_{,})_\alpha^T\Sigma_Y^{-1}\Sigma_X\Sigma_Y^{-1}(\mathcal{C}^{\mathrm{ind}} \bm{\lambda}_{,})_\beta.
    \label{eq:C_freq}
\end{align}

In \cref{eq:C_freq}, the statistical properties of the data assumed by choosing $\mathcal{L}_Y$ appears in $\Sigma_Y^{-1}$, as well as in the individual parameter covariance computed using $\mathcal{P}_Y$. Whereas the \textit{latent} statistical properties of data appears in $\Sigma_X$. The \cref{eq:C_freq} is an application of the \textit{delta method approximation} method for error propagation (\cite{Efron}, see their Eq. (2.8)) for the Gaussian case, but also a generalization since we take account of covariance between parameters and different \textit{latent} and \textit{analysis} data covariances respectively $\Sigma_X$ and $\Sigma_Y$. We can explore the solution when, in this formalism, we get $\mathcal{C}^{\mathrm{ens}}=\mathcal{C}^{\mathrm{ind}}=\mathcal{C}$, i.e. when the covariance of the estimator when considering an infinite realisation of the data $\bm{N}$ is consistent with the individual parameter covariance obtained from a single dataset $\bm{N}$. Following the Gaussian approximation of individual posteriors in the parameter space, we can use the Fisher forecast defined in Eq.~\eqref{eq:Fisher_formula_gaussian} for $\mathcal{C}^{\mathrm{ind}}$, since the Eq. \eqref{eq:C_freq} is evaluated at $\bm{N} = \bm{\lambda}_0$. Using this, we can simplify the problem by trying to find the relation between the data covariances $\Sigma_X$ and $\Sigma_Y$ to show that

\begin{align}
    \label{eq:C_1}
    \mathcal{C}_{\alpha\beta} &= (\mathcal{C} \bm{\lambda}_{,})_\alpha^T\Sigma^{-1}(\mathcal{C} \bm{\lambda}_{,})_\beta\\
\label{eq:C_2}
    \mathcal{C}_{\alpha\beta}^{-1}&=\bm{\lambda}_{,\alpha}^T \Sigma^{-1}_X\bm{\lambda}_{,\beta}
\end{align}

where we defined the modified covariance matrix $\Sigma^{-1} =\Sigma_Y^{-1} \Sigma_X \Sigma_Y^{-1}$ in \cref{eq:C_1}, and \cref{eq:C_2} is given by Fisher forecasting the individual parameter covariance in \cref{eq:Fisher_formula_gaussian}, that is always true for Gaussian and Poisson likelihoods. Multiplying \cref{eq:C_1} by $\mathcal{C}_{\eta\alpha}^{-1}$ and summing over $\alpha$, we get

\begin{equation}
    \delta^{K}_{\eta\beta} = \bm{\lambda}_{,\eta}^T \Sigma^{-1} (\mathcal{C} \bm{\lambda}_{,})_\beta = \bm{\lambda}_{,\eta}^T \Sigma^{-1}\bm{\lambda}_{,\delta} \mathcal{C}_{\delta\beta}.
\end{equation}

By re-multiplying the above equation by $\mathcal{C}^{-1}_{\alpha\beta}$, we get the result
\begin{equation}
    \mathcal{C}_{\eta\alpha}^{-1}=\bm{\lambda}_{,\eta}^T \Sigma^{-1}\bm{\lambda}_{,\alpha},
    \label{eq:C_end}
\end{equation}
which is the same as \cref{eq:C_2} when setting $\Sigma_X=\Sigma_Y$. So we found that for the ensemble parameter covariance to be equal to the individual parameter covariance, this formalism requires that the data covariance of the analysis and latent likelihoods are the same. In this appendix, we considered only the first order in the Taylor expansion in \cref{eq:e1}. When posteriors are not Gaussian, this approximation is not valid. 

\subsection{Poisson analysis likelihood}
\label{app:var_est_2}

In the previous section, we found that the ensemble parameter and individual parameter covariances are equal when $\Sigma_X = \Sigma_Y$, when the analysis likelihood $\mathcal{L}_Y$ is a multivariate Gaussian distribution defined in \cref{eq:binned_gaussian_likelihood} with data covariance $\Sigma_Y$, and where the latent likelihood is not specified, but with data covariance matrix $\Sigma_X$.
In this section, we consider the case where the analysis likelihood $\mathcal{L}_Y$ is the Poissonian as defined in \cref{eq:poisson_likelihood}. To compute the ensemble forecast, we first need to evaluate $[\nabla \widehat{\theta}_\alpha]_i$ in \cref{eq:nabla_theta}. First, we have that the Poisson log-likelihood is,

\begin{equation}
    \ln \mathcal{L}_Y(\bm{N}|\bm{\theta}) = -\lambda_{\rm tot}(\bm{\theta}) + \sum_{k=1}^c \bm{N}_k\ln \lambda_k(\bm{\theta}) - \ln (\bm{N}_k!),
    \label{eq:lnL_poisson}
\end{equation}
where $\lambda_{\rm tot}(\bm{\theta})$ is the total predicted abundance over all the mass-redshift bins. As given in \cref{eq:nabla_lnL}, we take the first derivative of the log-likelihood with respect to the data vector component $\bm{N}_i$, giving
\begin{equation}
    \nabla_i \ln \mathcal{L}_Y(\bm{N}|\bm{\theta}) = \ln \lambda_i(\bm{\theta}) - \psi(\bm{N}_i),
\end{equation}
where the $\psi$ is known as the digamma function and is defined as
\begin{equation}
    \psi(\bm{N}_i) =  -\gamma + \sum_{k'=1}^{\bm{N}_i} \frac{1}{k'},
\end{equation}
where $\gamma \approx 0.577$ is the Euler–Mascheroni constant, and the sum on the right side is the $m$-th harmonic number. We note that the quantity $\ln (\bm{\lambda}_0)_i-\psi(\bm{N}_i)$ is independent of $\bm{\theta}$, so it can be removed from the calculation of the covariance in \cref{eq:nabla_theta}. $[\nabla \widehat{\theta}_\alpha]_i$ now verifies the general form
\begin{equation}
    [\nabla \widehat{\theta}_\alpha]_i= \mathrm{Cov}_{\mathcal{P}_Y}[\{\theta_\alpha, \ln\lambda_i(\bm{\theta)}\}|\bm{N}].
\end{equation}
Next, taking the first order of the Taylor expansion of $\lambda_i(\bm{\theta})$ with respect to $\bm{\theta}$, we get $\lambda_i(\bm{\theta}) = (\lambda_0)_i + (\lambda_{,\gamma})_i(\bm{\theta}-\bm{\theta}_0)_\gamma$. Using $\ln (1 + x) \approx x + o(x^2)$, we get finally that
\begin{equation}
    [\nabla \widehat{\theta}_\alpha]_i = \frac{1}{(\bm{\lambda}_0)_i} \mathcal{C}^{\rm ind}_{\alpha \gamma}(\lambda_{,\gamma})_i = [\Sigma_Y^{-1}]_{ij}\mathcal{C}^{\rm ind}_{\alpha \gamma}(\lambda_{,\gamma})_j,
\end{equation}
where $\Sigma_Y = \mathrm{diag}[\{(\bm{\lambda}_0)_i\}_{1\leq i \leq c}]$ is the covariance matrix of the data considering the Poisson distribution in \cref{eq:poisson_likelihood} (in that case, abundances are not correlated and the variance of each $k$-th cluster count is the shot noise $(\bm{\lambda}_0)_k$), and $\mathcal{C}_{\alpha \gamma}^{\rm ind} = \mathrm{Cov}_{\mathcal{P}_Y}[\{\theta_\alpha, \theta_\beta\}|\bm{\lambda}_0]$ is the posterior covariance matrix as defined in \cref{eq:individual_covariance}. We find the same expression as for the Gaussian case in \cref{eq:nabla_theta_2}. To test the condition $\mathcal{C}^{\rm ens}=\mathcal{C}^{\rm ind}$ for the Poisson analysis likelihood, we first need to compute the fisher forecast for the Poisson case. From the Poisson likelihood in \cref{eq:lnL_poisson}, we have that the second derivative of the Poisson log-likelihood writes
\begin{equation}
    (\ln \mathcal{L})_{,\alpha\beta} = -(\lambda_{\rm tot})_{,\alpha\beta} + \sum_{k=1}\widehat{N}_k\left(\frac{(\lambda_{,\alpha\beta})_k}{\lambda_k} - \frac{(\lambda_{,\alpha})_k(\lambda_{,\beta})_k}{\lambda^2_k}\right),
\end{equation}
Following the Fisher matrix general formula in \cref{eq:Fisher_formula} and using the equality
\begin{equation}
    (\ln \mathcal{L})_{,\alpha\beta} = \frac{\mathcal{L}_{,\alpha\beta}}{\mathcal{L}} - (\ln \mathcal{L})_{,\alpha}(\ln \mathcal{L})_{,\beta},
\end{equation}
we get after few calculations the Fisher information matrix for the Poisson likelihood that writes
\begin{equation}
\mathrm{F}_{\alpha\beta} = \sum_{k=1}^{c}\frac{(\lambda_{,\alpha})_k(\lambda_{,\beta})_k}{(\lambda_0)_k} = \bm{\lambda}_{,\alpha}^T \Sigma^{-1}_Y\bm{\lambda}_{,\beta},
\end{equation}
where $\mathcal{C}^{\rm ind}$ can be forecasted as $\mathrm{F}^{-1}$. We see that the forecasted ensemble and individual parameter covariances for the Poisson analysis likelihood are obtained using the diagonal data covariance $\Sigma_Y$ with only shot noise in \cref{eq:Fisher_formula_gaussian} for $\mathcal{C}^{\rm ind}$ and in \cref{eq:cov_real_formula} for $\mathcal{C}^{\rm ens}$. As shown in \cref{app:var_est_1}, the equality $\mathcal{C}^{\rm ind} = \mathcal{C}^{\rm ens} = \mathcal{C}$ is then verified when $\Sigma_X = \Sigma_Y$ (result obtained following the derivation from \cref{eq:C_1} to \cref{eq:C_end}).

\
\bsp	
\label{lastpage}
\end{document}